\documentclass[preprint]{aastex}

\usepackage{amssymb}
\usepackage{lscape}

\newcommand\rsun{\hbox{\,R$_\odot$}}

\def\meas{ {\mbox{\tiny measured}}}
\def\true{ {\mbox{\tiny true}}}

\shorttitle{Data Analysis: Radio/Optical Interferometry}
\shortauthors{Monnier \& Allen}

\begin{document}



\title{Radio \& Optical Interferometry:\\[0.1in]
Basic Observing Techniques and Data Analysis \\[0.2in]
 \textit{To be published by Springer in Volume 2 of \\
 Planets, Stars, \& Stellar Systems}
}
\author{
J.D. Monnier\altaffilmark{1} \& {R.J.~Allen\altaffilmark{2}
}
}
\altaffiltext{1}{monnier@umich.edu; Univ. Michigan Astronomy Dept., 941 Dennison Bldg, Ann Arbor, MI 48109-1090, USA.}
\altaffiltext{2}{rjallen@stsci.edu; Space Telescope Science Institute, 3700 San Martin Drive, Baltimore, MD 21218, USA}
\email{}





\begin{abstract} 
  Astronomers usually need the highest angular resolution possible
  when observing celestial objects, but the blurring effect of
  diffraction imposes a fundamental limit on the image quality from
  any single telescope.  Interferometry allows light collected at
  widely-separated telescopes to be combined in order to synthesize an
  aperture much larger than an individual telescope thereby improving
  angular resolution by orders of magnitude.  Because diffraction has
  the largest effect for long wavelengths, radio and millimeter wave
  astronomers depend on interferometry to achieve image quality on par
  with conventional large-aperture visible and infrared telescopes.
  Interferometers at visible and infrared wavelengths extend angular
  resolution below the milli-arcsecond level to open up unique
  research areas in imaging stellar surfaces and circumstellar
  environments.

  In this chapter the basic principles of interferometry are reviewed
  with an emphasis on the common features for radio and optical
  observing.  While many techniques are common to interferometers
  of all wavelengths, crucial differences are identified that will
  help new practitioners to avoid unnecessary confusion and common
  pitfalls.  The concepts essential for writing observing proposals
  and for planning observations are described, depending on the
  science wavelength, the angular resolution, and the field of view
  required.  Atmospheric and ionospheric turbulence degrades the
  longest-baseline observations by significantly reducing the
  stability of interference fringes. Such instabilities represent a
  persistent challenge, and the basic techniques of phase-referencing
  and phase closure have been developed to deal with them.  Synthesis
  imaging with large observing datasets has become a routine and
  straightforward process at radio observatories, but remains
  challenging for optical facilities.  In this context the
  commonly-used image reconstruction algorithms CLEAN and MEM are
  presented.  Lastly, a concise overview of current facilities
  is included as an appendix.

\end{abstract}

\section{Interferometry in Astronomy}
\label{sec:introduction}
\subsection{Introduction}

The technique of interferometry is an indispensable tool for modern
astronomy.  Typically the telescope diameter $D$ limits the angular
resolution for an imaging system to $\Theta\approx\frac{\lambda}{D}$
owing to diffraction, but interferometry allows the achievement of
angular resolutions $\Theta\approx \frac{\lambda}{B}$ where the
baseline $B$ is set by the distance between telescopes.
Interferometry has permitted the angular resolution at radio
wavelengths to initially reach, and now to significantly surpass, the
resolution available with both ground- and space-based optical
telescopes. Indeed, radio astronomers routinely create high-quality
images with high sensitivity, high angular resolution, and a large
field-of-view using arrays of telescopes such as the Very Large Array
(VLA), the Combined Array for Research in Millimeter-wave Astronomy
(CARMA), and now the Atacama Large Millimeter Array
(ALMA). Interferometer arrays are now the instruments of choice for
imaging the wide range of spatial structures found both for Galactic
and for extragalactic targets at radio wavelengths.

At optical wavelengths, interferometry can improve the angular
resolution down to the milli-arcsecond level, an order-of-magnitude
better than even the Hubble Space Telescope. While atmospheric
turbulence limits the sensitivity much more dramatically than for the
radio, optical interferometers can nevertheless measure the angular
sizes of tens of thousands of nearby Galactic objects and even a
growing sample of distant Active Galactic Nuclei (AGN).  Recently,
optical synthesis imaging of complex objects has been demonstrated
with modern arrays of 4--6 telescopes, producing exciting results and
opening new avenues for research.

Both radio and optical interferometers also excel at precision
astrometry, with the potential for {\em micro-arcsecond}-level
precision for some applications.  Currently, it is ground-based radio
interferometry (e.g.\ the VLBA) that provides the highest astrometric
performance, although ground-based near-IR interferometers are
improving and measure different astronomical phenomena.

This chapter will provide an overview of interferometry theory and
present some practical guidelines for planning observations and for
carrying out data analysis at the premier ground-based radio and
optical interferometer facilities currently available for research in
astronomy.  In this chapter,  the term ``radio'' will be used as a shorthand
for the whole class of systems from sub-mm to decametric wavelengths
which usually employ coherent high-frequency signal amplification,
superheterodyne signal conversion, and digital signal processing,
although detailed instrumentation can vary substantially.  Likewise,
the term ``optical'' generally describes systems employing
direct detection, i.e. the direct combination of the signals from each
collector without amplification or mixing with locally-generated
signals\footnote{It should be emphasized, however, that this
  distinction is somewhat artificial; the first meter-wave radio
  interferometers $\approx 65$ years ago were simple Michelson adding
  interferometers employing direct detection without coherent
  high-frequency signal amplification. At the other extreme,
  superheterodyne systems are currently routinely used at wavelengths
  as short as $10\mu$m, such as the UC Berkeley ISI facility.}.

Historically, radio and optical interferometry have usually been
discussed and reviewed independently from each other, leaving the
student with the impression that there is something fundamentally
different between the two regimes of wavelength. Here a
different approach is taken, presenting a unified and more
wavelength-independent view of interferometry, nonetheless noting
important practical differences along the way.  This perspective
will demystify some of the differing terminology and techniques in a more
natural way, and hopefully will be more approachable for
a broad readership seeking general knowledge.  For a more much
detailed treatment of radio interferometry specifically, refer to 
the classic text by \citet{tms2001} and the series of lectures in the
NRAO Summer School on Synthesis Imaging \cite[e.g.][]{taylor1999}.
Optical interferometry basics have been covered in individual reviews by
\citet{quirrenbach2001} and by \citet{monnier2003}, and a useful
collection of course notes can be found in the NASA-Michelson Course
Notes \citep{mss2000} and ESO-VLTI summer school proceedings
\citep{malbet2007}.  Recently, a few textbooks have been published on
the topic of optical interferometry specifically, including
\citet{labeyrie2006}, \citet{glindemann2010}, and
\citet{saha2011}. Further technical details can also be found in
Chapters 7 and 13 of the first volume in this series.

This chapter begins with a 
brief history of interferometry and its scientific impact on
astronomy, a basic scientific context for newcomers
that illustrates why the need for better angular resolution has been
and continues to be one of the most important drivers for technical
innovation in astronomy.

\subsection{Scientific impact}
Using interferometers in a synthesis imaging array allows designers
 to decouple the diffraction-limited angular resolution of a
telescope (which improves linearly with the telescope size) from its
collecting area (which, for a filled aperture, grows quadratically
with the size).  In the middle of the 20th century, radio astronomers
faced a challenge in their new science; the newly-discovered ``radio
stars'' were bright enough to be observed with radio telescopes of
modest collecting area, but the resolution of conventional ``filled
aperture'' reflecting telescopes was woefully inadequate (by one or
more orders of magnitude) to measure the positions and angular sizes
of these enigmatic new cosmological objects with a precision
sufficient to permit an identification with an optical object. Thus
separated-element interferometry, although first applied in astronomy
at optical wavelengths \citep{michelson1921}, began to be applied in
the radio with revolutionary results.

\begin{figure}[hbt]
\begin{center}
\includegraphics[angle=0,clip,width=6in]{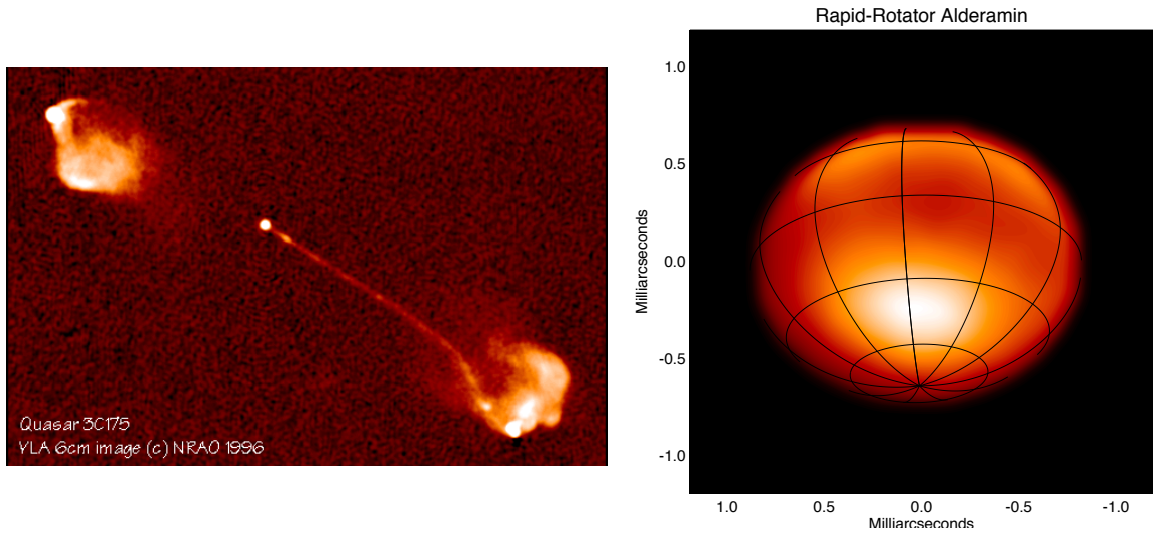} 
\figcaption{\footnotesize  Examples of imaging with interferometric arrays.
(left) Radio image of jets in quasar 3C175 produced by the Very Large Array
with field of view $\sim$1~arcminute$\sim$200 kpc and angular resolution
of 0.35~arcseconds
\citep[reprinted with permission][]{bridle1994}.  
(right) Near-infrared image of the rapidly-rotating star Alderamin produced
by the CHARA Array with
field of view of 7$\rsun$$\sim$2.5~milli-arcseconds and angular resolution
of 0.6~milliarcseconds
\citep[reprinted with permission][]{zhao2009}.  The hot polar region and the
cool equator 
is caused by the effect of ``gravity darkening.''
\label{fig_niceimages}}
\end{center}
\end{figure}

At radio wavelengths, the epoch of rapid technological development
began more than 50 years ago, and now interferometry is the
``workhorse'' technique of choice for most radio astronomers in the
world. A steady stream of exciting new results has flowed from these
instruments even until today, and a complete census of the major
discoveries to date would be very lengthy indeed. Here, our attention is 
focussed on the earliest historical discoveries that
\textit{required} radio interferometers, and we have listed our nominations in
Table \ref{table:discovery}.  In Figure~\ref{fig_niceimages}a,  the spectacular image of radio jets in
quasar 3C175 by the VLA is shown to illustrate the  high-fidelity imaging that is
possible using today's radio facilities.

\begin{deluxetable}{llll}
\tabletypesize{\scriptsize}
\tablecaption{Some historically-important astronomical results made possible by interferometry \label{table:discovery} }
\tablewidth{0pt}
\tablehead{ 
\colhead{Astronomical Result} & \colhead{Date} & \colhead{Facility} & 
\colhead{References\tablenotemark{a}} }
\startdata \hline
\multicolumn{4}{c}{\em Radio Interferometry\tablenotemark{b}} \\
\hline
Solar radio emission from sunspots  & 1945-46 & Australia, Sea cliff interferometer & R1 \\
First Radio Galaxies NGC 4486 \& NGC 5128 & 1948 & New Zealand, Sea cliff interferometer & R2 \\
Identification of Cygnus A& 1951-53 & Cambridge, W\"urzburg antennas & R3 \\
Cygnus A double structure & 1953 & Jodrell Bank, Intensity interferometer & R4 \\ 
AGN superluminal motions & 1971 & Haystack-Goldstone VLBI & R5 \\ 
Dark matter in spiral galaxies & 1972-78 & Caltech interferometer, Westerbork SRT & R6 \\
Spiral arm structure \& kinematics & 1973-80 & Westerbork SRT & R7 \\
Compact source in Galactic center & 1974 & NRAO Interferometer & R8 \\ 
Gravitational lenses & 1979 & Jodrell Bank Mk1 + Mk2 VLBI & R9 \\ 
NGC 4258 black hole & 1995 & NRAO VLBA & R10 \\ 
\hline
\multicolumn{4}{c}{\em Optical Interferometry} \\
\hline
Physical diameters of hot stars  & 1974 & Narrabri Intensity Interferometer& O1 \\
Empirical effective temperature scale for giants & 1987 & I2T/CERGA & O2 \\
Survey of IR Dust Shells     & 1994 & ISI & O3 \\
Geometry of Be star disks & 1997 & Mark III & O4 \\ 
Near-IR Sizes of YSO disks & 2001 & IOTA & O5 \\ 
Pulsating Cepheid $\zeta$~Gem & 2001 & PTI & O6 \\ 
Crystalline silicates in inner YSO disks  & 2004 & VLTI & O7 \\ 
Vega is a rapid rotator  & 2006 & NPOI  & O8 \\ 
Imaging gravity-darkening on Altair & 2007 & CHARA & O9 \\
Near-IR sizes of AGN & 2009 & Keck-I & O10 
\enddata
\tablenotetext{a}{References: 
R1: \citet{pawsey1946,mccready1947}.
R2: \citet{bolton1949}. 
R3: \citet{smith51,baade1954}.
R4: \citet{JennisonGupta1953}.
R5: \citet{WhitneyEtAl71,CohenEtAl71}.
R6: \citet{RogstadShostak72,Bosma81a,Bosma81b}.
R7: \citet{AllenEtAl73,RotsShane75,Rots75,Visser80a,Visser80b}.
R8: \citet{GossEtAl03}.
R9: \citet{PorcasEtAl79,WalshEtAl79}.
R10: \citet{MiyoshiEtAl95}. \\
O1: \citet{hb1974}.
O2: \citet{dibenedetto1987}.
O3: \citet{danchi1994}.
O4: \citet{quirrenbach1997}.
O5: \citet{millangabet2001}.
O6: \citet{lane2001}.
O7: \citet{vanboekel2004}.
O8: \citet{peterson2006}.
O9: \citet{monnier2007}.
O10: \citet{kishimoto2009}.
}
\tablenotetext{b}{Radio list in part from \citet{WilkinsonEtAl04} and R.D. Ekers (2010, priv. comm.), with additions by one of the authors (RJA). Historical material prior to 1954 is also from W.M. Goss (2011, private communication) and \citet{Sullivan09}.
}
\end{deluxetable}

Modern long-baseline optical interferometry started approximately 30
years after radio interferometry, following the pioneering experiments
and important scientific results with the Narrabri intensity
interferometry \citep[e.g.,][]{hb1974} and the heterodyne work of the
Townes' group at Berkeley \citep{jbt1974}.  The first successful
direct interference of stellar light beams from separated telescopes
was achieved in 1974 \citep{labeyrie1975} and this was followed by
about twenty years of two-telescope (i.e., single baseline)
experiments which measured the angular diameters of a variety of
objects for the first time.  The first imaging arrays with more than
two telescopes were constructed in the 1990s, and the COAST
interferometer was first to make a true optical synthesis image using
techniques familiar to radio astronomers \citep{baldwin1996}.  Keck
and VLT interferometers both include 8-m class telescopes, making them
the most sensitive facilities in the world.  Recently, the CHARA array
has produced a large number of new images in the infrared using
combinations of 4 telescopes simultaneously.
Table~\ref{table:discovery} lists a few major scientific
accomplishments in the history of optical interferometry showing the
diversity of contributions in many areas of stellar astronomy and even
recent extragalactic observations of active galactic nuclei.  With
technical and algorithm advances, model-independent imaging has become
more powerful and a state-of-the-art image from the CHARA array is presented in 
Figure~\ref{fig_niceimages}b, showing
the surface of the rapidly rotating star Alderamin.

\section{Interferometry in theory and practice}

 \subsection{Introduction}

 The most basic interferometer used in observational astronomy
 consists of two telescopes configured to observe the same object and
 connected together as a Michelson interferometer.  Photons 
 collected at each telescope are brought together to a central
 location and combined coherently at the ``correlator''
 (radio term) or the ``combiner'' (optical term). For wavelengths
 longer than $\sim$0.2 millimeters, the free-space electric field is
 usually converted into cabled electrical signals and coherently
 amplified at the focus of each telescope. The celestial signal
 is then mixed with a local oscillator signal sent to both telescopes
 from a central location, and the difference frequency transmitted in
 cables back to the centrally-located correlator. For shorter
 wavelengths, cable losses increase, and signal transmission
 moves eventually into free space in a more "optical" mode, using
 mirrors and long-distance transmission of light beams in (sometimes
 evacuated) pipes.

 Depending on the geometry, the light from an astronomical object will
 in general be received at one telescope before it arrives at the
 other. If the fractional signal bandwidth $\Delta \nu$ is very narrow
 (either because of the intrinsic emission properties of the source,
 e.g. a spectral line, or because of imposed bandwidth limitations in
 amplifiers and/or filters), then the signal has a high degree of
 ``temporal coherence'', which is to say that the wave packet
 describing all the photons in the signal is extended in time by $\tau
 \approx 1/\Delta \nu$ seconds. Expressing the bandwidth in terms of
 wavelength,  $\Delta \nu = (c/\lambda_{0}^{2}) \cdot \Delta
 \lambda$ where $\lambda_0$ is the band center and $c$ is the speed of light.  
 The coherence time is
 then $\tau = (1/c) \cdot (\lambda_{0}^{2} / \Delta \lambda)$, and $c
 \cdot \tau$ is a scale size of the wave packet called the
 \textit{coherence length}, $L_c = c \cdot \tau = \lambda_{0}^{2} /
 \Delta \lambda$, \cite[e.g. ][Ch.\ 7]{hecht2002}. If the path
 difference between the two collectors in an interferometer is a
 significant fraction of $L_c$,  an additional time
 delay must be introduced, otherwise the fringe amplitude will decrease or even
 disappear. For ground-based systems, the geometry is continually
 changing for all directions in the sky (except in the directions to
 the equatorial poles), requiring a continually-changing additional
 delay to maintain the temporal coherence. The special location on
 the sky where the adjusted time delay is matched perfectly is
 often called the ``phase center'' or point of zero optical path delay (OPD), 
although such a condition actually
 defines the locus of a plane passing through the mid-point between
 the collectors and perpendicular to the baseline, and cutting the
 celestial sphere in a great circle. Since the telescope optics
 usually limits the field of view to only a tiny portion of this great
 circle, adjusting the phase center is the equivalent of ``pointing''
 the interferometer at a given object within that field of view.

\begin{figure}[hbt]
\begin{center}
\includegraphics[angle=0,width=4in]{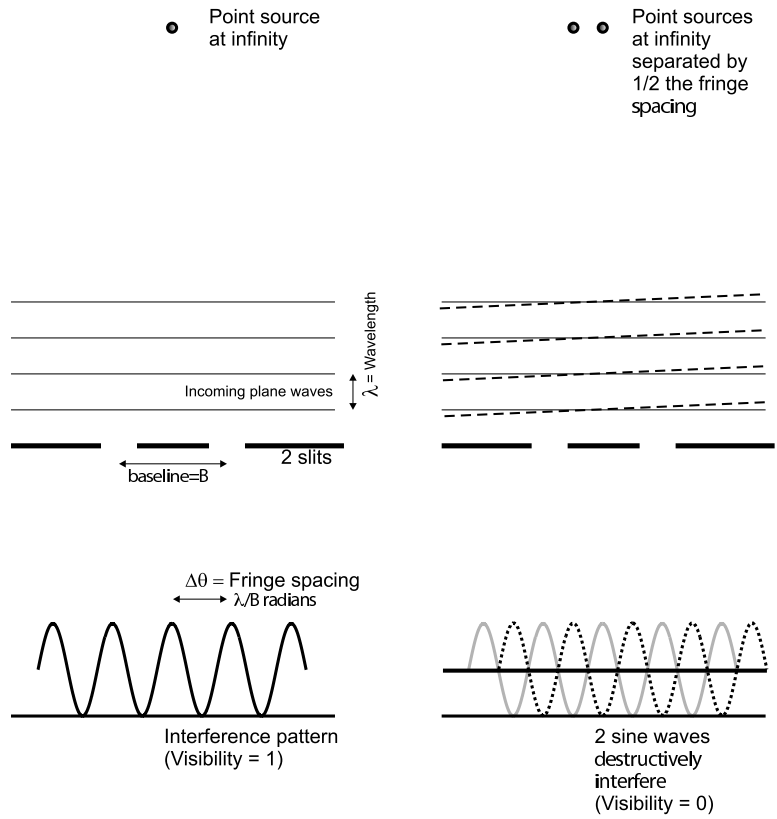} 
\figcaption{\footnotesize Basic operating principle
behind interferometry as illustrated by
Young's two-slit experiment \citep[adapted from][]{monnier2003}. 
\label{fig_schematic}}
\end{center}
\end{figure}

 The final step is to interfere the two beams to measure the
 \textit{spatial coherence} (often called the mutual coherence) of the
 electric field as sampled by the two telescopes. If the object
 observed is much smaller than the angular resolution of the
 interferometer, then interference is complete and one observes 100\%
 coherence at the correlator/combiner. However, objects that are {\em
   resolved} (i.e., much larger than the angular resolution of the
 interferometer) will show less coherence due to the fact that
 different patches of emission on the object do not interfere at the
 same time through our system.  Figure~\ref{fig_schematic} shows two
 simple cases of an interferometer as a Young's two-slit experiment to
 illustrate basic principles. At the left,  the interferometer is 
 made up of two slits and the response for a monochromatic point
 source (i.e., incoming plane waves) is shown.  The result should be familiar:
 an interference fringe modulating the intensity from 100\% to 0\%
 with a periodicity that corresponds to a fringe spacing of
 $\frac{\lambda}{B}$ on the sky.  Next to this panel is shown an example
 of two equal-brightness point sources separated by $\frac{1}{2}
 \frac{\lambda}{B}$, half the fringe spacing.  The location of
 constructive interference for one point coincides with the location
 of destructive interference for the other source. Since the two
 sources are mutually incoherent, the superposition of the two fringe
 results in an even light distribution, i.e. no fringe at all!  In
 optical interferometry language, the first example fringe has a
 fringe contrast (or visibility) of $1$ while the second example fringe
 has a visibility of $0$.

\begin{figure}[hbt]
\begin{center}
\includegraphics[width=6in]{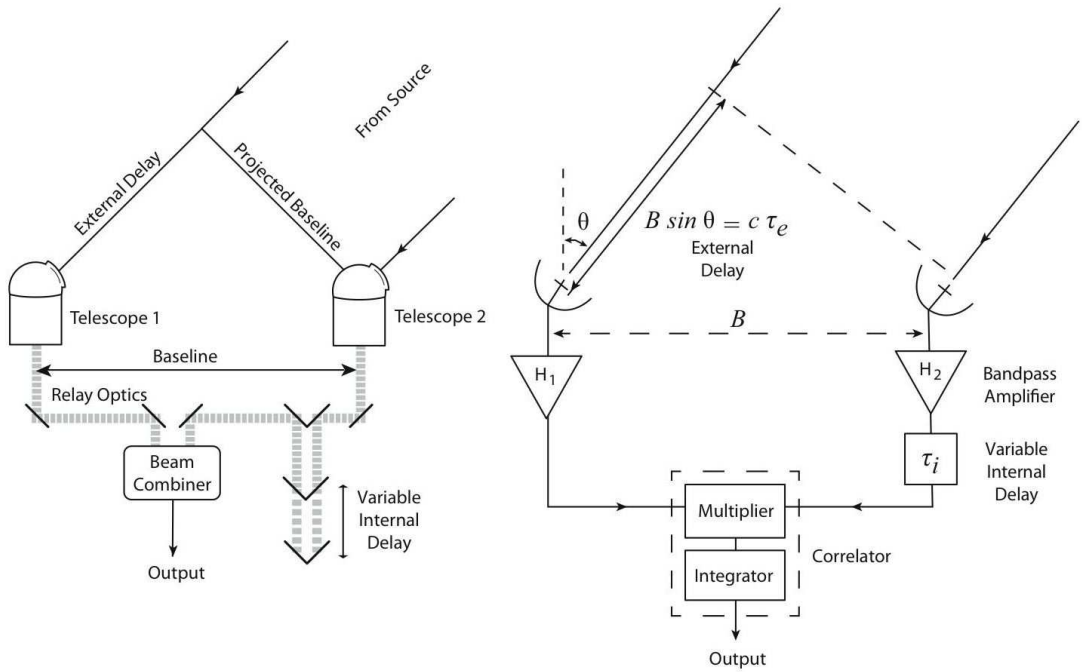}
\figcaption{\footnotesize  Here are more realistic examples of
long-baseline interferometers, both optical and radio, including
the light collectors and delay line.
Left panel: Optical interferometer, adapted from \citet{monnier2003}.
Right panel: Radio interferometer, adapted from Figure 2.3 in \citet{tms2001}.
See text for further discussion.  
\label{fig_interferometers}}
\end{center}
\end{figure}

 Figure~\ref{fig_interferometers} contains a schematic of a basic
 interferometer as typically realized for both radio and optical
 configurations.  While instrumental details vary immensely in how one
 transmits and interferes the signals for radio, millimeter, infrared, and
 visible-light interferometers, the basic principles are the same.
 The foundational theory common to all interferometers will be
 introduced next.

 \subsection{Interferometry in theory}
\label{vcz}
The fundamental equation of interferometry is typically derived by
introducing the van-Cittert Zernike Theorem and a complete 
treatment can be found in Chapter~3 of the book by \citet{tms2001}.  Here
the main result will  be presented without proof, 
beginning by defining an interferometric
observable called the {\em complex visibility},
$\tilde{{\mathcal{V}}}$.  The visibility can be derived from the
intensity distribution on the sky $I(\vec{\sigma})$ using a given
interferometer baseline $\vec{B}$ (which is the separation vector
between two telescopes) and the observing wavelength $\lambda$:
\begin{equation}
\label{eq:vcz}
\tilde{\mathcal{V}} = |\mathcal{V}| e^{i \phi_{\mathcal{V}}} = \int_{\rm sky} A_N (\vec{\sigma}) I(\vec{\sigma})   e^{-\frac{2\pi i}{\lambda} \vec{B}\cdot\vec{\sigma}} d\Omega
\end{equation}
Here, the $\vec{\sigma}$ represents the vector pointing from the
center of the field-of-view (called the ``phase center'') to a given
location on the celestial sphere using local (East, North) equatorial
coordinates and the telescope separation vector $\vec{B}$ also using
east and north coordinates.  The modulus of the complex visibility
$|\mathcal{V}|$ is referred to as the {\em fringe amplitude or
  visibility} while the argument $\phi_{\mathcal{V}}$ is the {\em
  fringe phase}. $A_N(\vec{\sigma})$ represents the normalized pattern
that quantifies how off-axis signals are attenuated as they are
received by a given antenna or telescope.  In this treatment the astronomical 
object is assumed to be small in angular size in order to 
ignore the curvature of the celestial sphere.

The physical baseline $\vec{B}$ can be decomposed into components $\vec{u}
= (u,v)$ in units of observing wavelength along the east and north
directions (respectively) as projected in the direction of our target.
The vector $\vec{\sigma}=(l,m) $ also can be represented in rectilinear
coordinates on the celestial sphere, where $l$ points along local east
and $m$ points north\footnote{There are several different coordinate
  systems in use to describe the geometry of ground-based
  interferometers used in observing the celestial sphere; \cite[see
  e.g.][Chapter 4 and Appendix 4.1]{tms2001}.}.  Here, $l$ and $m$
both have units of radians.  Equation \ref{eq:vcz} now becomes:
\begin{equation}
\label{eq:vcz2}
\tilde{\mathcal{V}}(u,v) = |\mathcal{V}| e^{i \phi_{\mathcal{V}}} = \int_{l,m} A_N (l,m) I(l,m)   e^{-2\pi i (u l + v m)} dl dm
\end{equation}
%
%
The fundamental insight from Equation~\ref{eq:vcz2} is that an
interferometer is a Fourier Transform machine -- it converts an
intensity distribution $I(l,m)$ into measurements of Fourier components
$\tilde{\mathcal{V}}(u,v)$ for all the baselines in the array represented
by the $(u,v)$ coverage.  Since an intensity distribution can be
described fully in either image space or Fourier space, the collection
of sufficient Fourier components using interferometry allows for an
image reconstruction through an inverse Fourier Transform process, although practical limitations lead to compromises in the quality of such images.

 \subsection{Interferometry in practice}
\label{interferometry_practice}

In this section, the similarities and differences between radio and
optical interferometers are summarized along with the reasons for the main
differences.  Interested readers can find more detailed on specific hardware implementations in Volume I of this series.

Modern radio and optical interferometers typically use conventional
steerable telescopes to collect photons from the target.  In the
radio, a telescope is often called an antenna; it is typically a
parabolic reflector with a very short focal length (f/D $\approx 0.35$
is common), with signal collection and initial amplification
electronics located at the prime focus. Owing to the large value of
$\Delta\Theta\sim\frac{\lambda}{{\rm Diameter}}$, the diffraction
pattern of the antenna aperture is physically a relatively large
region at the prime focus. This fact, coupled with the cost and
complexity of duplicating and operating many low-noise receivers in
close proximity to each other, has meant that antennas used in radio
astronomy typically have only a ``single-pixel'' signal collection
system (a dipole or a ``feed horn'') at the prime focus\footnote{Very
  recently, several radio observatories have begun to equip their
  antennas (and at least one entire synthesis telescope) with arrays
  of such feeds.}. Light arriving from various directions on the sky
are attenuated depending on the shape of the diffraction pattern,
written as $A_N$ in Equation \ref{eq:vcz2} and often called the
``antenna pattern'' or the ``primary beam''. The signal collection
system may be further limited to a single polarization mode, although
systems are common that simultaneously accept both linear (or both
circular) polarization states. After initial amplification, the signal
is usually mixed with a local oscillator to ``down-convert'' the high
frequencies to lower frequencies that can more easily be amplified and
processed further.  These lower-frequency signals from the separate
telescopes can also be more easily transported over large distances to
a common location using e.g.\ coaxial cable, or by modulating an
optical laser and using fiber optics. This common location houses the
``back end'' of the receiver, where the final steps in signal analysis
are carried out including band definition, correlation, digitizing,
and spectral analysis. In some cases, the telescope signals are
recorded onto magnetic media and correlated at a later time and in a
distant location (e.g.\ the ``Very Long Baseline Array'' or global
VLBI).

In the optical, the light from the object is generally focused by the
telescope, re-collimated into a compressed beam for free-space
transport, and then sent to a central location in a pipe which is
typically evacuated to avoid introducing extra air dispersion and
ground-level turbulence.  In rare cases, the light at the telescope is
focused directly into a single-mode fiber, which is the dielectric
equivalent to the metallic waveguides used in radio and millimeter
receivers.  Note that atmospheric seeing is very problematic for even small
visible and infrared telescopes while it is usually negligible compared
to the diffraction limit for even the largest radio and mm-wave telescopes.

Both radio and optical interferometers must delay the signals from
some telescopes to match the optical paths. After mixing to a lower frequency, radio interferometers can use switchable lengths of coaxial cable in order
to introduce delays.  More recently, the electric fields can be
directly digitized with bandwidths of $>5$ GHz, and these ``bits'' can
be saved in physical memory and then recalled at a later time.  For
visible and infrared systems, switchable fiber optics are not
practical due to losses and glass dispersion; the only solution is to use
an optical ``free-space'' delay line consisting of a retroreflector moving
on a long track and stabilized through laser metrology to compensate for
air path disturbances and vibrations in the building.

In a radio interferometer, once all the appropriate delays have been
introduced the signals from each telescope can be combined.  Early
radio signal correlators operated in an ``optical'' mode as simple
adding interferometers, running the sum of the signals from the two
arms through a square-law detector. The output of such a detector
contains the product of the two signals. Unfortunately, the desired
correlation product also comes with a large total power component
caused by temporally-uncorrelated noise photons contributed
(primarily) by the front-end amplifiers in each arm of the
interferometer plus atmosphere and ground noise.  This large signal
demanded excellent DC stability in the subsequent electronics, and it
was not long in the history of radio interferometry before engineers
found clever switching and signal-combination techniques to suppress
the DC component. These days signal combiners deliver only the product
of the signals from each arm, and are usually called
``correlators''\footnote{This is not all advantageous; if the data is
  intended to be used in an imaging synthesis, the absence of the
  total power component means that the value of the map made from the
  data will integrate to zero. In other words, without further
  processing the image will be sitting on a slightly-negative
  ``floor''. If more interferometer spacings around zero are also
  missing, the floor becomes a ``bowl''. All this is colloquially
  called ``the short-spacing problem'', and it adversely affects the
  photometric accuracy of the image. A significant part of the
  computer processing ``bag of tricks'' used to ``restore'' such
  images is intended to address this problem, although the only proper
  way to do that is to obtain the missing data and incorporate it into
  the synthesis.}. Most modern radio/millimeter arrays use digital
correlators that introduce time lags between all pairs of telescopes
in order to do a full temporal cross-correlation. This allows a
detailed wavelength-dependent visibility to be measured, i.e., an
interferometric spectrum with $R=\frac{\lambda}{\Delta\lambda}>100000$
if necessary.  By most metrics, radio correlators have reached their
fundamental limit in terms of extracting full spectral and spatial
information and can be fairly sophisticated and complex to configure
when correlating multiple bandpasses simultaneously with high spectral
resolution\footnote{At millimeter and sub-millimeter wavelengths,
  correlators still do not attain the maximum useful bandwidths for
  continuum observations}.

In the visible and infrared, the electric fields can not be further
amplified without strongly degrading the signal-to-noise ratio, and so
parsimonious beam combining strategies are common that split the
signal using e.g.\ partly-reflecting mirrors into a small number of
pairs or triplets.  Furthermore, most optical systems have only modest
spectral resolutions of typically $R\sim40$ in order to maintain
high signal-to-noise ratio, although a few specialized
instruments exist that reach $R>1000$ or even $R>30000$. Signal
combination finally takes place simply by mixing the light beams
together and modulating the relative optical path difference, either
using spatial or temporal encoding.  The total power measurement in a
visible-light or infrared detector will reveal the interference fringe
and a Fourier analysis can be used to extract the complex visibility
$\tilde{\mathcal{V}}$.

Because the ways of measuring visibilities are quite different, radio
and optical interferometrists typically report results in different
units.  Radio/mm interferometers measure correlated flux density in
units of Jansky ($10^{-26}$ W\, m$^{-2}$\, Hz$^{-1}$), just as suggested by
Equation \ref{eq:vcz2}\footnote{Recall that an integration of specific intensity over solid angle results in a flux density, often expressed in Jansky.}.  
In the optical however, interferometers
tend to always measure a normalized visibility that varies from $0$ to $1$ --
this is simply the correlated signal normalized by the total power.
One can convert the latter to correlated
flux density by simply multiplying by the known total flux density of
the target at the observed wavelengths, or otherwise by carrying out a
calibration of the system by a target of known flux
density.

\subsubsection{Quantum limits of amplifiers}
\label{quantumnoise}
The primary reason why radio and optical interferometers differ so
much in their detection scheme is because coherent amplifiers would introduce
too much extraneous noise at the high frequencies encountered in the optical
and infrared. This difference is fundamental and is explored in more
detail in this section.

At radio frequencies there are huge numbers of photons in a typical
sample of the electromagnetic field, so the net phase of a packet of
radio photons (either from the source or from a noisy receiver) is
well-defined and amplifiers can operate coherently. The ultimate
limits which apply to such amplifiers are dictated by the uncertainty
principle as stated by Heisenberg. Beginning with the basic ``position
- uncertainty'' relation $\Delta x \; \Delta p_x \geq h/4 \pi$, it is
easy to derive the ``energy - time'' relation $\Delta E \; \Delta t
\geq h/4\pi$. Since the uncertainty in the energy of the $n$ photons
in a wave packet can be written as $\Delta E = h \nu \; \Delta n$ and
the uncertainty in the phase of the aggregate as $\Delta \phi = 2 \pi
\nu \; \Delta t$, this leads to the equivalent uncertainty relation
$\Delta \phi \; \Delta n \geq 1/2$. 

An ideal amplifier which adds no noise to the input photon stream
leads to a contradiction of the uncertainty principle. The following
argument shows how this happens \citep[adapted from][]{Heffner1962}:
Consider an ideal
coherent amplifier of gain $G$ which creates new photons in phase
coherence with the input photons, and assume it adds no incoherent
photons of its own to the output photon stream. With $n_1$ photons
going into such an amplifier, there will be $n_2 = Gn_1$ photons at the
output, all with the same phase uncertainty $\Delta \phi_2 = \Delta
\phi_1$ with which they went in. In addition, in this model it is expected that
 $\Delta n_2 = G \Delta n_1$ (no additional ``noise'' photons 
unrelated to the signal). But according to the same uncertainty relation,
the photon stream coming out of the amplifier must also satisfy
$\Delta \phi_2 \; \Delta n_2 \geq 1/2$. This would imply that $\Delta
\phi_1 \; \Delta n_1 \geq \frac{1}{2G}$, which for large $G$ says that
 the input photon number and wave packet phase could be measured with
essentially no noise. But this contradicts the same uncertainty
relation for the input photon stream, which requires that $\Delta
\phi_1 \; \Delta n_1 \geq 1/2$. This contradiction shows that one or
more of our assumptions must be wrong. The argument can be saved if 
the amplifier itself is required to add noise of its own to the photon
stream; the following heuristic construction shows how.  Using the
identity $\Delta n_2 = (G-N) \cdot \Delta n_1 + N\Delta n_1$ at the
output (where $N$ is an integer $N \geq 1$), and referring this noise
power back to the input by dividing it with the amplifier gain $G$, this leads to
 $(1 - N/G) \cdot \Delta n_1 + (N/G) \cdot \Delta n_1$ at the input to the
amplifier, which for large $G$ is $\Delta n_1$. The smallest possible
value of N is $1$.  This preserves the uncertainty relation at the
expense of an added minimum noise power of $ h \nu$ at the
input. \citet{Oliver1965} has elaborated and generalized this argument
to include all wavelength regimes, and has shown that the minimum
total noise power spectral density $\psi_\nu$ of an ideal amplifier
(relative to the input) is
\begin{equation}
\psi_\nu = \frac{h \nu}{e^{(h \nu /k T)} - 1} + h \nu \ \ \mbox{Watts/Hz ,}
\end{equation}
where T is the kinetic temperature that the amplifier input faces in
the propagation mode to which the amplifier is sensitive. For $h \nu <
kT$ this reduces to $\psi_\nu\approx kT$ Watts/Hz, which can be
called the "thermal" regime of radio astronomy. For $h \nu > kT$ this
becomes $\psi_\nu \approx h \nu$ Watts/Hz in the "quantum" regime of
optical  astronomy. The crossover point where the two contributions are
equal is where $h \nu / kT = \ln{2}$, or at $\lambda_c \cdot T_c =
20.75$ (mm~K). As an illustration of the use of this equation, consider
this example: The sensitivity of high-gain radio-frequency amplifiers
can usually be improved by reducing their thermodynamic temperatures.
However, for instance at a wavelength of 1 mm, it might be unnecessary (depending on details of the signal chain) to aim
for a high-gain amplifier design to lower the thermodynamic temperature
below about 20K, since at that point the sensitivity is in any case limited
by quantum noise. At even shorter wavelengths, the rationale for cooled
amplifiers disappears, and at optical wavelengths amplifiers are clearly
not useful since the noise is totally dominated by spontaneous
emission\footnote{Although amplifiers are currently used in the long-distance
transmission of near-IR (digital) communication signals in optical fibers,
the signal
levels are relatively large and low noise is not an important
requirement.} and is equivalent to thermal emission temperatures of thousands
of degrees. The extremely faint signals common in modern optical
observational astronomy translate into very low photon rates, and the
addition of such irrelevant photons into the data stream by an amplifier
would not be helpful.

\subsection{Atmospheric Turbulence}
\label{turbulence}
So far, the analysis of interferometer performance has assumed a
perfect atmosphere.  However, the electromagnetic signals from cosmic
sources are distorted as they pass through the intervening media on
the way to the telescopes. These distortions occur first in the
interstellar medium, followed by the interplanetary medium in the
solar system, then the Earth's ionosphere, and finally the Earth's
lower atmosphere (the troposphere) extending from an altitude of 
$\approx 11$ km down to ground level. The media involved in the first
three sources of distortion contain ionized gas and magnetic fields,
and their effects on signal propagation depend strongly on wavelength
(generally as $\propto \lambda^2$) and polarization. At wavelengths
shorter than about 10 cm the troposphere begins to
dominate. Molecules in the troposphere (especially water vapor) become
increasingly troublesome at frequencies above 30 GHz (1 cm wavelength), and the atmosphere is essentially opaque beyond 300 GHz except for two rather narrow (and not very clear) ``windows'' from 650-700 and 800-900  GHz which are usable only at the highest-altitude sites. The next atmospheric windows appear in the IR at wavelengths less than about 15 microns. The optical window opens around one micron, and closes again for wavelengths shorter than about 350 nm.

The behavior of the troposphere is thus of prime importance to
ground-based astronomy at wavelengths from the decimeter-radio to the
optical. Interferometers are used in the study of structure in the
troposphere, and a summary of approaches and results with many
additional references is given in \citet[][Ch.~13]{tms2001,carilli1999,sutton1996}.
A discussion oriented towards optical wavelengths can be found in
\citet{quirrenbach2000}. Since the main focus here is on using
interferometers to measure the properties of the cosmic sources
themselves,  our discussion is limited to some ``rules of thumb''
for choosing the interferometer baseline length and the time interval
between measurements of the source and of a calibrator in order to
minimize the deleterious effects of propagation on the fringe
amplitudes and (especially) fringe phases.

\subsubsection{Phase fluctuations -- length scale}

Owing to random changes in the refractive index of the atmosphere and
the size distribution of these inhomogeneities, the path length for
photons will be different along different parallel lines of
sight. This fluctuating path length difference grows almost linearly
with the separation $d$ of the two lines of sight for separations up
to some maximum, called the outer scale length (typically tens to hundreds of
meters, with some weak wavelength dependence), and is roughly
constant beyond that. Surprisingly, in spite of the differences in the
underlying physical processes causing refraction, variations in the
index of refraction are quite smooth across the visible and all the way
through to the radio. At short radio wavelengths, the fluctuations are dominated by turbulence in the water vapor content; at optical/IR
wavelengths, it is temperature and density fluctuations in dry air that dominate.

Using a model of fully-developed isotropic Kolmogorov turbulence for the 
Earth's atmosphere, the rms path length difference grows according to
$\sigma_d \propto d^{5/6}$ for a path separation $d$
\citep[see][Ch.~13, for references]{tms2001}. High altitude sites show
smaller path length differences as the remaining vertical thickness of
the water vapor layer decreases. Relatively large seasonal and diurnal
variations also exist at high mountain sites as the atmospheric
temperature inversion layer generally rises during the summer and
further peaks during mid-day. Variations in $\sigma_d$ by factors of
$\sim 10$ are not unusual \citep[see][Fig.~13.13]{tms2001}, but a
rough average value for a good observing site is
$\sigma_d \approx 1$ mm for
baselines $d \approx 1$ km at millimeter wavelengths, and
$\sigma_d \approx 1$ micron for baselines $d \approx 50$ cm at
infrared wavelengths.

The length scale fluctuations translate into fringe phase fluctuations
of $\sigma_\phi = 2\pi\sigma_d/\lambda$ in radians. The {\em maximum
  coherent baseline} $d_0$ is defined as that baseline length for
which the rms phase fluctuations reach 1 radian. Using the expressions
in the previous paragraph and coefficients suitable for the radio and 
optical ranges at the better observing sites, two useful approximations are $d_0 \approx 140 \cdot
\lambda^{6/5}$ meters for $\lambda$ in millimeters (useful at
millimeter radio wavelengths), and $d_0 \approx 10 \cdot
\lambda^{6/5}$ centimeters for $\lambda$ in microns (useful at IR
wavelengths). These two expressions are in fact quite similar; using the ``millimeter expression'' to calculate $d_0$ in the IR underestimates the value obtained from the ``IR expression'' by a factor of 2.8, which is at the level of precision to be expected.

At shorter wavelengths (visible and near-infrared), atmospheric
turbulence limits even the image quality of small telescopes.  This
has led to a slightly different perspective for the length scale that
characterizes atmospheric turbulence, although it is closely related to
the previous description.  The Fried length $r_0$ \citep{fried1965} is
the equivalent-sized telescope diameter whose diffraction limit matches
the image quality through the atmosphere due to {\em seeing}.
It turns out that this quantity is proportional to the length
scale where the rms phase error over the telescope aperture is
$\approx 1$ radian. In other words, 
apertures with diameters small compared to $r_0$
are approximately diffraction limited, while larger apertures have 
resolution limited by turbulence to $\approx \lambda / r_0$. It
can be shown that, for an atmosphere with fully-developed Kolmogorov
turbulence, $r_0 \approx 3.2  d_0$ \citep[][Ch. 13]{tms2001}.

\subsubsection{Phase fluctuations -- time scale}

Although fluctuations of order one radian may be no more than a
nuisance at centimeter wavelengths, requiring occasional phase
calibration (see \S\ref{phasereferencing}), 
they will be devastating at IR and visible wavelengths
owing to their rapid variations in time. In order to relate the
temporal behavior of the turbulence to its spatial structure, a 
model of the latter is required along with some assumption for how that structure
moves over the surface of the Earth. One specific set of assumptions
is described in \citet[][Ch.~13]{tms2001}; however, for the purposes here
it is sufficient to use Taylor's ``frozen atmosphere'' model 
with a nominally-static phase screen that 
moves across the Earth's surface with the wind at speed $v_s$. This
phase screen traverses the interferometer baseline $d$ in a time $\tau_d = d
/ v_s$, at the conclusion of which the total path length variation is
$\sigma_d$. Taking the critical time scale $\tau_c$ to be when the rms
phase error reaches 1 radian, then $\tau_c \approx d_0 / v_s$ with
$d_0$ given in the previous paragraph. As an example consider a wind
speed of 10 m/s; this leads to $\tau_c \approx 14$ seconds at
$\lambda = 1$ mm, and $\approx 10$ milliseconds at $\lambda = 1$ micron.
Clearly the techniques required to manage these variations will be very
different at the two different wavelength regimes, even though the
magnitude of the path length fluctuations (in radians of phase) are similar. Representative
values of these quantities are collected in Table \ref{table:fluctuations}.

\subsubsection{Calibration -- Isoplanatic Angle}

The routine calibration of interferometer phase and amplitude is
usually done by observing a source with known position and intensity
inter-leaved in time with the target of interest. At centimeter
wavelengths and longer, the discussion in the previous section
indicates that such measurements can be done on time scales of minutes
to hours, providing ample time to re-position telescopes elsewhere on
the sky in order to observe a calibrator. But how close to the target
of interest does such a calibrator have to be? Ideally, the calibrator
ought to be sufficiently nearby on the celestial sphere that the line
of sight traverses a part of the atmosphere with substantially the
same phase delay as the line of sight to the target. This angle is
called the \textit{isoplanatic angle} $\Theta_{iso}$; it characterizes
the angular scale size over which different parts of the incoming
wavefront from the target encounter closely similar phase shifts,
thereby minimizing the image distortion. The isoplanatic angle can be
roughly estimated by calculating the angle subtended by an $r_0$-sized
patch at a height $h$ that is characteristic for the main source of
turbulence; hence, roughly $\Theta_{iso}\approx\frac{r_0}{h}$.  Within
a patch on the sky with this angle, the telescope/interferometer PSF
remains substantially constant, retaining the convolution relation
between the source brightness distribution and the image. Some
approximate values are given in Table \ref{table:fluctuations} as a
guide.

At visible and near-IR wavelengths, Table \ref{table:fluctuations}
shows that the isoplanatic angle is very small, smaller than an
arcminute.  Unfortunately, the chance of having a suitably bright and
point-like object within this small patch of the sky is very low.
Even if an object did exist, it would be nearly impossible to
repetitively re-position the telescope and delay line at the
milli-second level timescale needed to ``freeze'' the turbulence
between target and calibrator measurements.  Special techniques to
deal with this problem will be discussed further in section
\ref{phasereferencing}.

\begin{deluxetable}{ccccc}
\tabletypesize{\scriptsize}
\tablecaption{Approximate baseline length, Fried length, and time scales for a 1-radian rms phase fluctuation in the Earth's troposphere and a wind speed of 10 m/s.\tablenotemark{a} \label{table:fluctuations} }
\tablewidth{0pt}
\tablehead{ 
 \colhead{} & \colhead{Max. Coherent} & \colhead{Fried length} & \colhead{Time scale} &\colhead{Isoplanatic angle at zenith} \\
  \colhead{Wavelength} & \colhead{Baseline $d_0$} & \colhead{$r_0$} & \colhead{$\tau_c$} & \colhead{$\Theta_{\rm iso}$}
}
\startdata
0.5 $\mu$m (visible) & 4.4 cm & 14 cm & 4.4 ms & $5.5''$ \\ \hline
2.2 $\mu$m (near-IR) & 26 cm & 83 cm & 26 ms & $33''$ \\  \hline
1 mm (millimeter) & 140 m & 450 m & 14 sec & $3.5^{\circ}$ \\ \hline
10 cm (radio) & 35 km & 112 km & 58 min & large\tablenotemark{b} \\ \hline
\enddata

\tablenotetext{a}{~From parameters for Kolmogorov turbulence given in \citet[][Ch.~13]{tms2001}, and in \citet[][Table 2]{Woolf82}. The inner and outer scale lengths are presumed to remain constant in these rough approximations. Values are appropriate for a good observing site and improve at higher altitudes.  See \S\ref{turbulence} for more discussion.}

\tablenotetext{b}{~Limited in practice by observing constraints such as telescope slew rates and elevation limits, and source availability.}
\end{deluxetable}

\section{Planning Interferometer Observations}
\label{observing}
The issues to consider when writing an interferometer observing
proposal or planning the observations themselves include: the desired
sensitivity (i.e., the unit telescope collecting area, the number of
telescopes to combine at once, the amount of observing time), the
required field-of-view and angular resolution (i.e.,the shortest and
longest baselines), calibration strategy and expected systematic
errors (i.e., choosing phase and amplitude calibrators), the expected
complexity in the image (i.e., the completeness of u,v coverage, do
science goals demand model-fitting or model-independent imaging), and
the spectral resolution (i.e., correlator settings, choice of combiner
instrument).  Many of these issues are intertwined, and the burden on
the aspiring observer to reach a compatible set of parameters can be
considerable. Prospective observers planning to use the VLA are
fortunate to have a wide variety of software planning tools and user's
guides already at their disposal, but those hoping to use more
experimental facilities or equipment which is still in the early
phases of commissioning will find their task more challenging.

Here, the most common issues encountered during interferometer
observations will be introduced. In many ways this is more of a list
of things to worry about rather than a compendium of solutions. The
basic equations and considerations have been collected in
Table~\ref{table_planning}.  In order to obtain the latest advice on
optimizing a request for observing time, or to plan an observing run,
observers ought to consult the web sites, software tools, and human
assistants available for them at each installation (see Appendix for
a list of current facilities).

\begin{deluxetable}{ll}
\tabletypesize{\scriptsize}
\tablecaption{Planning Interferometer Observations
\label{table_planning}}
\tablewidth{0pt}
\tablehead{ 
 \colhead{Consideration} & \colhead{Equation} 
}
\startdata
Angular Resolution & $\Theta = \frac{1}{2} \frac{\lambda}{B_{\rm max}}$\\
\hline
Spectral Resolution & $R=\frac{\lambda}{\Delta \lambda} = \frac{c}{\Delta v}$ \\
\hline
Field-of-View & \\
\em primary beam & $\Delta\Theta \sim \frac{\lambda}{D_{\rm Telescope}}$\\
\em bandwidth-smearing & $\Delta\Theta \sim R \cdot \frac{\lambda}{B_{\rm max}}$ \\
\em time-smearing & $\Delta\Theta \sim \frac{230}{\Delta t_{\rm minutes}} \frac{\lambda}{B_{\rm max}}$ \\
\hline
Phase Referencing & \\
\em Coherence Time  & see Table~\ref{table:fluctuations}\\
\em Isoplanatic Angle  & see Table~\ref{table:fluctuations}\\
\hline
\enddata
\end{deluxetable}

\subsection{Sensitivity}
Fortunately modern astronomers can find detailed documentation on the
expected sensitivities for most radio and optical interferometers
currently available.  Indeed, the flexibility of modern
instrumentation sometimes defies a back-of-the-envelope estimation for
the true signal-to-noise ratio (SNR) expected for a given observation.
In order to better understand what limits sensitivity for 
real systems, the dominant noise sources and the key
parameters affecting signal strength are introduced.  Most of the focus will be for
observations of point sources since resolved sources do not contribute
signal to all baselines in an array and this case must be treated with some care.

Here, the discussions of the radio and optical cases 
are separate
because of the large differences in the nature of the noise processes
(e.g., see \S\ref{quantumnoise}) and the associated nomenclature.
Radio and optical observations lie at the two limits of Bose-Einstein
quantum statistics that govern photon arrival rates 
\citep[e.g.,][see \S6.3]{pathria1972}. At long wavelengths, the
occupation numbers are so high that the statistics evolve into the Gaussian
limit and where the root-mean-square (rms) fluctuation in the detected power $\Delta$P is
proportional to the total power P itself (e.g., $\Delta$Power $\propto$
Power).  On the other hand, in the optical limit, the sparse
occupation of photon states results in the familiar Poisson statistics
where the level of photon fluctuations $\Delta$N is proportional to
$\sqrt{N}$.  Most of the SNR considerations for interferometers are
in common with single-dish radio and standard optical photometry, and 
so interested readers are referred to the relevant chapters in
Volumes 1 and 2 of this series.


\subsubsection{Radio Sensitivity}
\label{radiosensitivity}

The signal power spectral density $P_\nu$ received by a radio telescope of
effective area $A_e$ (${\rm m}^2$) from a celestial point source of flux
density $S_\nu$ (Jansky = Watts/${\rm m}^2$/Hz) is $P_\nu = A_e \cdot S_\nu$
(Watts/Hz). It is common to express this as the power which would be
delivered to a radio circuit (wire, coaxial cable, or waveguide) by a
matched termination at a physical temperature $T_A$, called the ``antenna
temperature'', so that $T_A = A_eS_\nu/2k$ (Kelvin) where $k$ = Boltzmann's
constant and the factor 1/2 accounts for the fact that, although the
telescope's reflecting surface concentrates both states of polarization at a
focus, the ``feed'' collects the polarization states separately.
As described in section \ref{quantumnoise}, the amplifier which follows must
add noise; this additional noise power (along with small contributions from
other extraneous sources in the telescope field of view) $P^{\rm s}_\nu$ can
likewise be expressed as $P^{\rm s}_\nu = kT_s/2$, where  $T_s$ is
the ``system temperature.'' The rms fluctuations in this noise power will
limit the faintest signals that can be distinguished. As mentioned in the
previous paragraph, these fluctuations are directly proportional to the
receiver noise power itself, so $\Delta T_s \propto T_s$.
They will also be inversely proportional to the
square root of the number of samples of this noise present in the receiver
passband. The coherence time of a signal in a bandwidth $\Delta \nu$ is
proportional to $1/\Delta \nu$, so in an integration time $\tau$ there are of
order $\tau \Delta \nu$ independent samples of the noise, and
the statistical uncertainty will improve as $1/\sqrt{\tau\Delta \nu}$. The
ratio of the rms receiver noise power fluctuations to the signal power is
therefore:
\begin{equation}
\Delta T_s / T_A \propto \frac{2 k T_s}{A_e S \sqrt{\tau \Delta \nu}} \; .
\end{equation}
The minimum detectable signal $\Delta S$ is defined as the value of $S$ for
which this ratio is unity. For this ``minimum'' value of S the equation
becomes:
\begin{equation}
\Delta S = \frac{f_c \cdot k T_s}{A_e \sqrt{\tau \Delta \nu}} \; ,
\end{equation}
The coefficient of proportionality $f_c$ for this equation is of order unity,
but the precise value depends on a number of details of how the receiver 
operates. These details include whether the receiver output contains both 
polarization states, whether both the in-phase and the quadrature channels
of the complex fringe visibility are included, whether the receiver operates
in single- or double-sideband mode, and how precisely the noise is quantized
if a digital correlator is used. Further discussion of the various
possibilities is given in \citet[][Chapter 6]{tms2001}. For the present 
purpose, it suffices to notice that the sensitivity for a specific radio
interferometer system improves only slowly with integration time and with
further smoothing of the frequency (radial velocity) resolution. The most
effective improvements are made by lowering the system temperature and by
increasing the collecting area. 

The point-source sensitivity continues to improve as telescopes are added
to an array.  An array of $n$ identical telescopes contains $N_b =
n(n-1)/2$ distinct baselines. If the signals from each telescope are
split into multiple copies, $N_b$ interference pairs can be made. The
rms noise in the flux density on a point source including all the
data is then
\begin{equation}
\Delta S = \frac{f_c \cdot k T_s}{A_e\sqrt{N_b \tau \Delta\nu}} \; .
\end{equation}

So far the discussion has been made for isolated point sources.
Extended sources are physically characterized by their surface
brightness power spectral density $B_{\rm surf}(\alpha,\delta,\nu)$
(Jansky/steradian) and by the angular resolution of the observation 
as expressed by the solid angle $\Omega_b$ of the synthesized beam in steradians (see \S\ref{imaging}).
By analogy with the discussion of rms noise power from thermal sources
given earlier, it is usual to express the surface brightness power
spectral density for an extended sources  in terms of a temperature.
This conversion
of units to Kelvins is done using the Rayleigh-Jeans approximation to
the Planck black-body radiation law, although the radiation observed
in the image is only rarely thermally-generated. The conversion from
$B_{\rm surf}(\alpha,\delta,\nu)$ (Jansky/steradian) to $T_b$ in
Kelvins is
\begin{equation}
T_b = \frac{\lambda^2 B_{\rm surf}}{2 k \Omega_b} \; ,
\end{equation}
which requires ($h\nu/kT << 1$) if the radiation is thermal; otherwise,  this conversion can be viewed merely as a convenient change of units. The rms\ brightness temperature sensitivity in a radio synthesis image from receiver noise alone is then
\begin{equation}
\Delta T_b = \frac{f_c \lambda^2 T_s}{2 A_e \Omega_b \sqrt{N_b \tau \Delta \nu}} \; .
\end{equation}

The final equations above for the sensitivity on synthesis imaging
maps shows that the more elements one has, the better the flux density
sensitivity will be. For example if one compares an array of $N_b=20$
baselines with an array containing $N_b=10$ baselines, the flux
density SNR is improved by a factor $\sqrt{2}$ no matter where the
additional 10 baselines are located in the $u,v$ plane. However, the
brightness temperature sensitivity does depend critically on the
actual distribution of baselines used in the synthesis. For instance,
if the same number of telescopes is ``stretched out'' to double the
maximum extent on the ground, the equations above show that the flux
density sensitivity $\Delta S$ remains the same, but the brightness
temperature sensitivity $\Delta T_b$ is worse by a factor of 4 since
the synthesized beam is now 4~times smaller in solid angle. This is a
serious limitation for spectral line observations where the source of
interest is (at least partially) resolved and where the maximum surface brightness is modest. For instance, clouds of
atomic hydrogen in the Galactic ISM never seem to exceed surface
brightness temperatures of $\approx 80$ K, so the maximum achievable
angular resolution (and hence the maximum useable baseline in the
array) is limited by the receiver sensitivity. This can only be
improved by lowering the system temperature on each telescope or by
increasing the number of interferometer measurements with more
telescopes and/or more observing time.

A cautionary note is appropriate here. In the case of an optical image
of an extended object taken e.g.\ with charge-coupled device (CCD) camera on a 
filled aperture telescope, a simple way of improving the SNR is to
average neighboring pixels together thereby creating a smoothed image
of higher brightness sensitivity. At first sight, the equation for
$\Delta T_b$ above suggests that this should also happen with
synthesis images, but here the improvement is not as dramatic as it
may seem at first sight. The reason is that the action of smoothing is
equivalent to discarding the longer baselines in the $u,v$ plane; for
instance, reducing the longest baseline used in the synthesis by a
factor of 2 would indeed lead to an image with brightness temperature
sensitivity which is better by a factor of $2^2$, but the effective
reduction of the number of interferometers from $N$ to $N/2$ means
that the net improvement is only $2^{1.5}$. A better plan would have
been to retain all the interferometers but to shrink the array
diameter with the factor 2 by moving the telescopes into a more
compact configuration. This is one reason why interferometer arrays
are usually constructed to be reconfigurable.

\subsubsection{Visible and Infrared Sensitivity}
As mentioned earlier, the visible and infrared cases deviate
substantially from the radio case.  While the sensitivity is still
dependent on the collecting area of the telescopes ($A_e$), the
dominant noise processes behave quite differently.  In the visible and infrared (V/IR), noise
is generated by the random arrival times of the photons governed by
Poisson statistics $\Delta N = \sqrt{N}$, where $N$ is the mean number
of photons expected in a time interval $\tau$ and $\Delta N$ is the
rms variation in the actual measured number of photons.  Depending on
the observing setup (e.g., the observing wavelength, spectral
resolution, high visibility case or low visibility case),
 the dominant noise term can be Poisson noise from the
source itself, Poisson noise from possible background radiation, or
even detector noise.  Because of the centrality of Poisson
statistic, it is common to  work in units of total detected
photo-electrons $N$ within a time interval $\tau$, rather than
power spectral density $P_\nu$ or system temperature $T_S$. 
This conversion is straightforward:

\begin{eqnarray}
N & = & \eta \frac{P_\nu \Delta\nu }{h \nu} \tau \\
  & = & \eta \frac{S_\nu A_e \Delta\nu}{h \nu} \tau 
\end{eqnarray}
\noindent where $\eta$ represents the total system detection
efficiency which is the combination of optical transmission of system
and the quantum efficiency of the detector and the other variables are the same as for the radio case introduced in the last section.

For the optical interferometer, atmospheric turbulence limits the size
of the aperture that can be used without adaptive optics (the
atmosphere does not limit the useful size of the current generation of
single-dish mm-wave and radio telescopes).  The Fried parameter $r_0$
sets the coherence length and thus the max($A_e$)$\sim
r^2_0$. Likewise without corrective measures, the longest useful
integration time is limited to the atmospheric coherence time
$\tau\sim\tau_{\rm c}$.  There exists a {\em coherent volume} of
photons that can be used for interferometry, scaling like $r_0 \cdot
r_0 \cdot c \tau_c$. As an example, consider the coherent volume of
photons for decent seeing conditions in the visible ($r_0 \sim$
10\,cm, $\tau_c \sim$ 3.3\,ms). From this, the limiting magnitude can
be estimated by requiring at least 10 photons to be in this coherent
volume.  Assuming a bandwidth of 100~nm, 10 photons ($\lambda\sim$
550\,nm) in the above coherent volume corresponds to a V magnitude of
11.3, which is the best limit one could hope to achieve\footnote{Real
  interferometers will have a realistic limit about 1-2 orders of
  magnitude below the theoretical limit due to throughput losses and
  non-ideal effects such as loss of system visibility.}. This is more
than 14~magnitudes worse than faint sources observed by today's 8-m
class telescopes that can benefit from integration times measured in
hours instead of milli-seconds.  Because the atmospheric coherence
lengths and timescales behave approximately like
$\lambda^{\frac{6}{5}}$ for Kolmogorov turbulence, the coherent volume
$\propto \lambda^\frac{18}{5}$.  Until the deleterious atmospheric
effects can be neutralized, ground-based optical interferometers will
never compete with even small single-dish telescopes in raw
point-source sensitivity.

Under the best case the only source of noise is Poisson noise from the
object itself.  Indeed, this limit is nearly achieved with the best
visible-light detectors today that have read-noise of only a few
electrons.  More commonly, especially in the infrared, detectors
introduce the noise that limits sensitivity, typically 10-15~electrons
of read-noise in the near-IR for the short exposures required to
effectively freeze the atmospheric turbulence.  For wavelengths longer
than about 2.0$\mu$m (i.e., K, L, M, N bands), Poisson noise from the
thermal background begins to dominate over other sources of
noise. Highly-sensitive infrared interferometry will require a space
platform that will allow long coherence times and low thermal
background.  Please consult the observer manual for each specific
interferometer instrumentation to determine point-source sensitivity.

Another important issue to consider is that a low visibility fringe
($\mathcal{V}<<1$) is harder to detect than a strong one. Usually
fringe detection sets the limiting magnitude of an
interferometer/instrument, and this limit often scales like $N
\mathcal{V}$, the number of ``coherent'' photons. For readnoise or
background noise dominant situations (common in NIR), this means that
if the point-source ($\mathcal{V}=1$) limiting magnitude is 7.5 then a
source with $\mathcal{V}=0.1$ would need to be as bright as magnitude
5.0 to be detected.  The magnitude limit worsens even more quickly for low visibility fringes  when noise from the
source itself dominates, since brighter targets bring along greater noise.
Another common expression found in the literature is that the SNR for
a visible-light interferometer scales like $N \mathcal{V}^2$. This
latter result can be derived by assuming that the ``signal'' is the
average power spectrum $(N \mathcal{V})^2$ and the dominant noise
process is photon noise which has a power spectrum that scales like
$N$ here.

\subsubsection{Overcoming the Effects of  the Atmosphere: Phase Referencing, Adaptive Optics, and Fringe Tracking}
\label{phasereferencing}
As discussed above, the limiting magnitude will strongly
depend on the maximum coherent integration time that is set by the
atmosphere.  Indeed, this limitation is very dramatic, restricting
visible-light integrations to mere milli-seconds and millimeter radio
observations to a few dozen minutes.  For mm-wave and radio observations,
the large isoplanatic angle and long atmospheric coherence times allow
for real-time correction of atmospheric turbulence by using {\em phase
  referencing.}  

In a phase-referencing observing sequence, the telescopes in the array
will alternate between the (faint) science target and a (bright) phase
calibrator nearby in the sky.  If close enough in angle
(within the isoplanatic patch), then the turbulence will be the same
between the target and bright calibrator; thus, the high SNR
measurement of fringe phase on the calibrator can be used to account
for the atmospheric phase changes.  Another key aspect is that the
switching has to be fast enough that the atmospheric turbulence does
not change between the two pointings.  With today's highly-sensitive
radio and mm-wave receivers, enough bright targets exist to allow
nearly full sky coverage so that most faint radio source will have a
suitable phase calibrator nearby.\footnote{At the shortest sub-mm
  wavelengths, phase-referencing is quite difficult due to strong
  water vapor turbulence, but can be partially corrected using
  ``water-vapor monitoring'' techniques
  \citep[e.g.,][]{wiedner2001}.}  In essence, phase referencing means
that a fringe does not need to be detected within a single coherence
time $\tau_c$ but rather one can coherently integrate for as long as
necessary with sensitivity improving as $1/\sqrt{t}$.  In
\S\ref{dataanalysis}  a simple example is presented that demonstrates how
phase-referencing works with simulated data.

In the visible and infrared, phase referencing by alternate
target/calibrator sequences is practically impossible since $\tau_c <<
1$~second and $\Theta_{\rm iso}<< 1$~arcminute isoplanatic patch
size. In V/IR interferometry, observations still alternate between a
target and calibrator in order to calibrate the statistics of the
atmospheric turbulence but not for phase referencing.  A special case
exists for dual-star narrow-angle astrometry \citep{shao1992} where a
``Dual Star'' module located at each telescope can send light
from two nearby stars down two different beam trains to be interfered
simultaneously.  At K~band, the stars can be as far as
$\sim$30$\arcsec$ apart for true phase referencing.  This approach is
being attempted at the VLT \citep[PRIMA,][]{delplancke2006} and Keck
Interferometers \citep[ASTRA,][]{woillez2010}.  This technique can be
applied to only a small fraction of objects owing to the low sky density
of bright phase reference calibrators.

Adaptive optics (AO) can be used on large visible and infrared telescopes to
effectively increase the collecting area $A_e$ term in our signal
equation, allowing the full telescope aperture to be used for
interferometry.
AO on a 10-m class telescope potentially boosts infrared sensitivity
by $\times$100 over the seeing limit; however, this method still
requires a bright enough AO natural or laser guide star to operate.
Currently, only the VLT and Keck Interferometers have adaptive optics
implemented for regular use with interferometry.  A related technique
of {\em fringe tracking} is in more widespread use, whereby the
interferometer light is split into two channels so that light from one
channel is used exclusively for measuring the changing atmospheric
turbulence and driving active realtime path length compensation.  In
the meantime, the other channel is used for longer science
integrations (at VLTI, Keck, CHARA).  This method improves the
limiting magnitude of the system at some wavelengths if the object is
substantially brighter at the fringe tracking wavelength, such as for
dusty reddened stars. Fringe tracking sometimes can be used for very
high spectral observations of stars ordinarily too faint to observe at
high dispersion.

It is important to mention these other optical interferometer
subsystems (e.g., AO, fringe tracker) here because they are crucial
for improving sensitivity, but the additional complexities do pose a
challenge for observers.  Each subsystem has its own sensitivity
limit and now multiple wavelengths bands are needed to drive the
crucial subsystems. As an extreme example, consider the Keck
Interferometer Nuller \citep{colavita2009}.  The R-band light is used
for tip-tilt and adaptive optics, the H band is used to correct for
turbulence in air-filled Coude path, the K band is used to fringe
track and finally the 10$\mu$m light is used for the nulling work.  If
the object of interest fails to meet the sensitivity limit of any of
these subsystems then observations are not possible -- most strongly
affecting highly reddened sources like young stellar objects and
evolved stars.

\subsection{(u,v) Coverage}
\label{uvcoverage}

One central difference between interferometer and
conventional single-telescope observations is the concept of (u,v)
coverage.  Instead of making a direct ``image'' of the sky at the
focal plane of a  camera, the individual fringe visibilities for
each pair of telescopes are obtained.  As discussed in \S\ref{vcz},
each measured complex visibility is a single Fourier component of a
portion of the sky.  The goal of this subsection is to understand how
to estimate (u,v) coverage from the array geometry and which
characteristics of Fourier coverage affect the final reconstructed image.

\begin{figure}[hbt]
\begin{center}
\includegraphics[angle=0,height=3in]{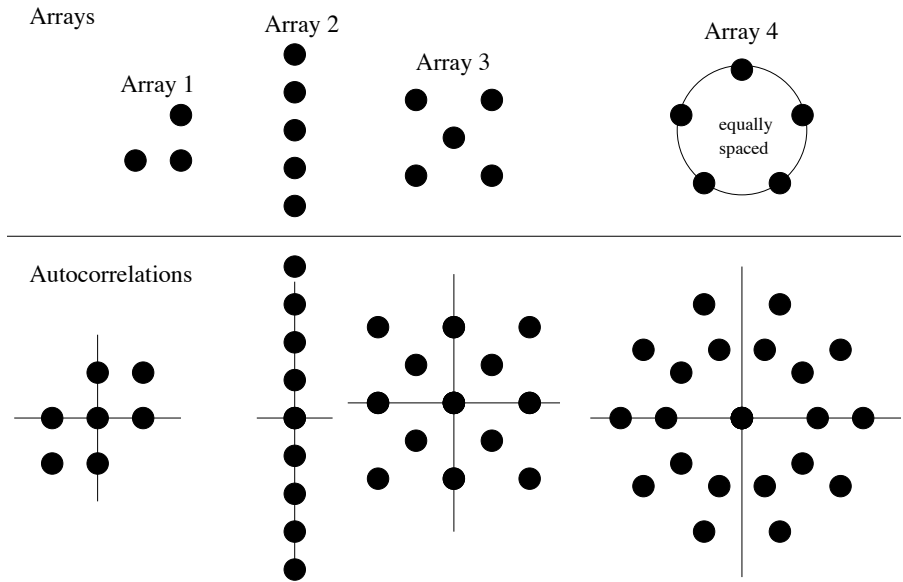}
\figcaption{\footnotesize Here the ``snapshot'' coverages of a few
simple interferometer
layouts are shown for an object located at zenith.  The top portion shows the
 physical layout of 4 examples arrays 
while the bottom portion shows the corresponding autocorrelation function,
which is the same as the (u,v) coverage.
\label{fig_uvcov1}}
\end{center}
\end{figure}

For a given layout of telescopes in an interferometer array, the
Fourier coefficients can be measured are determined by drawing
baselines between each telescope pair. To do this, an (x,y) coordinate
system is first constructed to describe the positions of each element
of the array; for ground-based arrays in the northern hemisphere, the
convention is to orient the +x axis towards the east and the +y axis
towards north.  The process of determining the complete ensemble of
(u,v) points provided by any given array can be laborious for arrays
with a large number of elements. A simple method of automating the
procedure is as follows. First, construct a distribution in the (x,y) plane
of delta functions of unit strength at the positions of all
elements. The (u,v) plane coverage can be obtained from the
two-dimensional autocorrelation of this distribution, as illustrated
in Figure~\ref{fig_uvcov1} for four simple layouts of array
elements. The delta functions for each array element are shown as dots
in the upper row of sketches in this figure, and the corresponding
dots in the u,v distributions are shown in the lower row of
autocorrelations.  Note that each point in the (u,v) plane is repeated
on the other side of the origin owing to symmetry; of course the
values of amplitude and phase measured on a source at one baseline
will be the same whether one thinks of the baseline as extending from
telescope 1 to telescope 2, or the converse. For an array of $N$
telescopes, one can measure ${{N}\choose{2}}=\frac{(N)(N-1)}{2}$
independent Fourier components.

Sometimes the array geometry may result in the (near-)duplication of
baselines in the (u,v) plane. This is the case for array \#2 in the
Figure~\ref{fig_uvcov1}, where the shortest spacing is duplicated 4
times, the next spacing is duplicated 3 times, the following spacing
is duplicated twice, and only the longest spacing of this array is
unique. While each of these interferometers does contribute
statistically independent data as far as the noise is concerned, it is
an inefficient use of hardware since the astrophysical information
obtained from such redundant baselines is essentially the same.  In
order to optimize the Fourier coverage for a limited number of
telescopes, a layout geometry should be {\em non-redundant}, with no
baseline appearing more than once, so that the maximum number  of Fourier components 
can be measured for a given array of telescopes. A
number of papers have been written on how to optimize the range and
uniformity of (u,v) coverage under different assumptions
\citep{golay1971, keto1997, holdaway1999}. Note that in the sketches
of Figure \ref{fig_uvcov1}, array \#4 provides superior coverage in
the u,v plane compared to arrays \#3 and \#2 with the same number of
array elements.

\begin{figure}[hbt]
\begin{center}
\includegraphics[angle=0,height=3in]{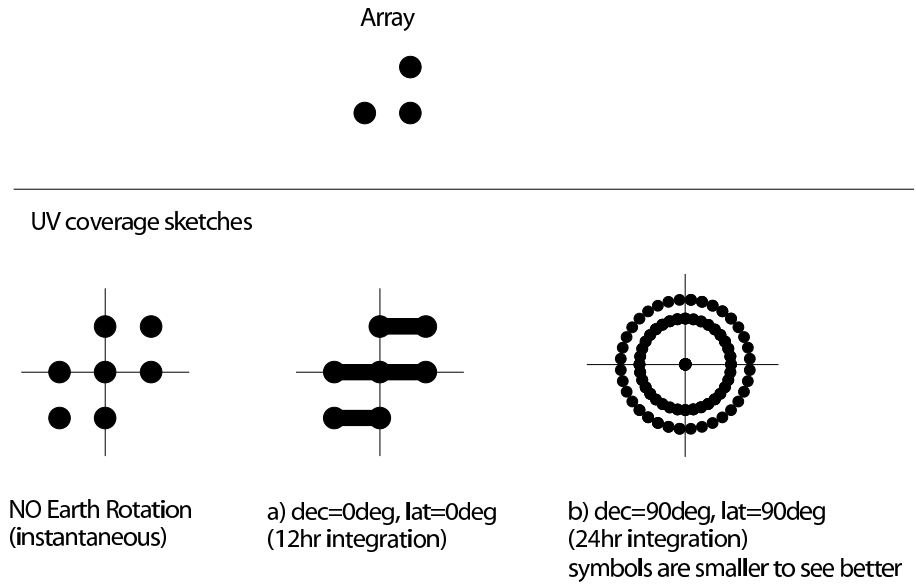}
\figcaption{\footnotesize The rotation of the Earth introduces changing
baseline projections as an object
traverses the sky. This figure take a simple 3-telescope array and shows
the ``(u,v) tracks'' for a few different examples of differing observatory
latitudes and target declinations.
\label{fig_uvcov2}}
\end{center}
\end{figure}

Finally note that the actual (u,v) coverage depends not on the {\em
  physical baseline separations} of the telescopes but on the {\em
  projected baseline separations} in the direction of the target. For
ground-based observing, a celestial object moves across the sky along
a line of constant declination, so the (u,v)-coverage is actually
constantly changing with time.  This is largely a benefit since earth
rotation dramatically increases the (u,v)-coverage without requiring
additional telescopes. This type of synthesis imaging is often called
{\em Earth rotation aperture synthesis}. The details depend on the
observatory latitude and the target declination, and a few simple
cases are presented in Figure~\ref{fig_uvcov2}.  In general, sources
with declinations very different from the local latitude will never
reach a high elevation in the sky, such that the north-south (u,v)
coverage will be foreshortened and the angular resolution in that
direction correspondingly reduced.

\begin{figure}[hbt]
\begin{center}
\includegraphics[clip,angle=0,width=6in]{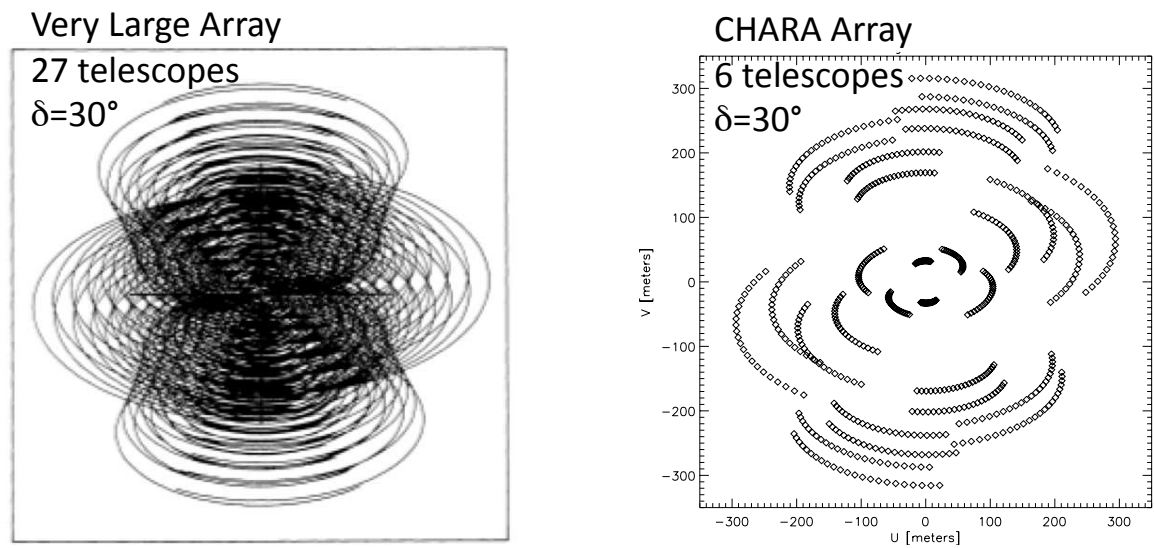}
\figcaption{\footnotesize This figure shows the dramatic advantage
in Fourier coverage
that the Very Large Array (VLA) possesses (left panel) compared to
the CHARA array (right panel).
The radio interferometer VLA, located in Socorro, NM, has 27 (movable)
elements and the coverage shown here corresponds to observations
$\pm$ 4.5 hours around source transit
\citep[adapted from][]{thompson1980}.  The optical interferometer
CHARA, located on Mt. Wilson, CA, 
has 6 (fixed) elements and
the coverage shown here corresponds to $\pm$3 hours from
transit \citep[adapted from][]{theo2005}.
The large gaps in Fourier coverage for CHARA limit its ability to
reconstruct highly complex images.
\label{fig_uvcov_compare}}
\end{center}
\end{figure}

Figure~\ref{fig_uvcov_compare} shows the actual Fourier coverage for
the 27-telescope Very Large Array (VLA) and for the 6-telescope CHARA
Array.  For $N=27$, the VLA can measure 351 Fourier components while
CHARA ($N=6$) can measure only 15 simultaneously. Notice also in this
figure that the ratio between the maximum baseline and the minimum
baseline is much larger for the VLA (factor of 50, A array) compared
to CHARA (factor of 10).

The properties of the (u,v)-coverage can be translated into some rough
characteristics of the final reconstructed image.  The final image
will have an angular resolution of $\sim\frac{\lambda}{B_{\rm max}}$,
and note that the angular resolution may not be the same in
all directions.  It is crucial to match the desired angular resolution
with the maximum baseline of the array because longer baselines will
over-resolve your target and have very poor (or non-existent)
signal-to-noise ratio (see discussion \S\ref{radiosensitivity}).  This
functionally reduces the array to a much smaller number of telescopes
which dramatically lowers both overall signal-to-noise ratio and the
ability to image complex structures.  For optical arrays that
combine only 3 or 4 telescopes, relatively few (u,v)
components are measured concurrently and this limits how much
complicated structure can be reconstructed\footnote{Fortunately,
  targets of optical interferometers are generally spatially compact
  and so sparser (u,v) coverage can often be  acceptable.}. From
basic information theory under best case conditions, one needs at least
as many independent visibility measurements as the number of
independent pixels in the final image. For instance, it will take
hundreds of components to image a star covered with dozens of spots of
various sizes, while only a few data points can be used to measure a
binary system with unresolved components.

\subsection{Field-of-view}

While the (u,v) coverage determines the angular resolution and quality of
image fidelity, the overall imaging field-of-view is constrained by a number
of other factors.

A common limitation for field-of-view is the primary beam pattern of
each individual telescope in the array and this was already discussed
in \S\ref{interferometry_practice}:
$\Delta\Theta\sim\frac{\lambda}{{\rm Diameter}}$.  This limit can be
addressed by {\em mosaicing}, which entails repeated observations over
a wide sky area by coordinating multiple telescope pointings within
the array and stitching the overlapping regions together into a single
wide-field image. This practice is most common in the mm-wave where
the shorter wavelengths result in a relatively small primary beam.  A
useful rule of thumb is that your field-of-view (in units of the
fringe spacing) is limited to the ratio of the baseline to the
telescope diameter.  Most radio and mm-wave imaging is limited by
their primary beam, however there is a major push to begin using
``array feeds'' to allow imaging in multiple primary beams
simultaneously.

Another limitation to field-of-view is the spectral resolution of the
correlator/combiner.  The spectral resolution of each channel can be
defined as $R=\frac{\lambda}{\Delta\lambda}$.  A combiner or
correlator can not detect a fringe that is outside the system
coherence envelope, which is simply related to the spectral resolution
$R$. The maximum observable field of view is $R$ times the finest
fringe spacing, or $\Delta\Theta \sim R \cdot \frac{\lambda}{B_{\rm
    max}}$, often referred to as the {\em bandwidth-smearing} limit.
Most optical interferometers and also Very Long Baseline
Interferometry (VLBI) are limited by bandwidth smearing.

A last limitation to field-of-view arises from temporal smearing of
data by integrating for too long during an observation. Because the
(u,v) coverage is constantly changing due to Earth rotation, time
averaging removes information in the (u,v)-plane resulting in reduced
field-of-view.  A crude field-of-view  limit based on this effect is $\Delta\Theta \sim
\frac{230}{\Delta t_{\rm minutes}} \frac{\lambda}{B_{\rm max}}$. Both
radio and V/IR interferometric data can be limited by temporal-smearing
if care is not taken in setting up the data collection, although this
limitation is generally avoidable.

\subsection{Spectroscopic Capabilities}

As for regular radio and optical astronomy, one tries to observe at
the crudest spectral resolution that is suitable for the science goal
in order to achieve maximum signal-to-noise ratio. However as just
discussed, spectral resolution does impact the imaging field-of-view, bringing in another dimension to preparations.  While each
instrument has unique capabilities that can not be easily generalized,
 most techniques will require dedicated spectral calibrations
as part of observing procedures.

``Spectro-interferometry'' is an exciting tool in radio and
(increasingly) optical interferometry. In this application, the
complex visibilities are measured in many spectral channels
simultaneously, often across a spectrally-resolved spectral line. This
allows the different velocity components to be imaged or modelled
independently.  For example, this technique can be used for observing
emitting molecules in a young stellar object to probe and quantify
Keplerian motion around the central mass or for mapping differential
rotation on the surface of a rotating star using photospheric
absorption lines \citep[e.g.,][]{kraus2008}.  Spectro-interferometry
is analogous to ``integral field spectroscopy'' on single aperture
telescopes, where each pixel in the image has a corresponding measured
spectrum. Another clever example of spectro-interferometry pertains to
maser sources in the radio: a single strong maser in one spectral
channel can be used as a phase calibrator for the rest of the spectral
channels \citep[e.g.,][]{greenhill1998}.

\section{Data Analysis Methods}
\label{dataanalysis}
After observations have been completed, the data must be
analyzed. Every  instrument will have a customized
software pipeline to take the recorded electrical signals and
transform into useful astronomical quantities.  That said, the
data reduction process is the similar for most systems and here the basic steps are outlined.

   \subsection{Data reduction  and calibration overview}
 
The goal of the data reduction is to produce calibrated complex
visibilities and related observables such as closure phases (see
\S\ref{closurephase}).  As discussed in \S\ref{observing}, the basic
paradigm for interferometric observing is to switch between
data of well-known system calibrators and the target of
interest.  This allows for calibration of changing atmospheric
conditions by monitoring the actual phase delay through the atmosphere
(in radio) or by statistically correcting for decoherence from
turbulence (in optical).

\begin{figure}[hbt]
\begin{center}
\includegraphics[angle=90,width=6in]{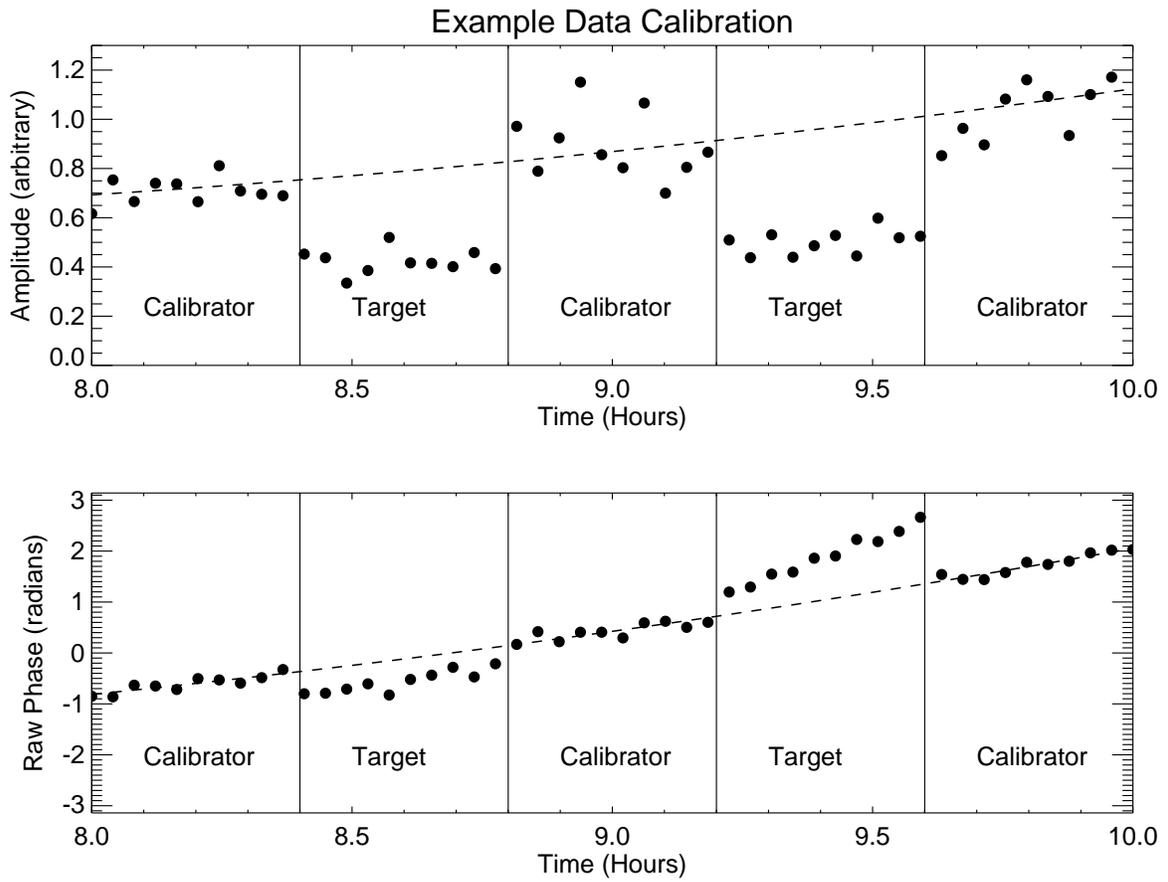}
\figcaption{\footnotesize First Step of Data Reduction
\label{calsrccal_a}}
\end{center}
\end{figure}

One begins by plotting the observed fringe amplitude versus time.
Figure~\ref{calsrccal_a} shows a schematic example of how data
reduction might proceed for the case of high quality radio
interferometry observations, such as taken with the EVLA.  Here 
the observed fringe amplitude and phase for a
calibrator-target-calibrator sequence is presented.  Notice that in this example
the fringe amplitude of the calibrator is drifting up with time, as is
the observed phase.  As long as the switching time between target and
calibrator is faster than instrumental gain drifts and atmospheric
piston shifts, a simple function can be fitted to the raw
calibrator measurements and then interpolated to produce the calibration
curves to for the target.  Here  a 2nd order polynomial has been used to
approximate the changing amplitude
and phase response.  This figure contains  an example for only a single
baseline, polarization, and spectral channel; there will be hundreds
or thousands of panels like this in a dataset taken with an instrument
as  the EVLA or ALMA.

The justification for this fitting procedure can be expressed mathematically.
As the wave traverses the atmosphere, telescope, and interferometer
beamtrain/waveguides, the electric field can have its phase and
amplitude modified.\footnote{In general, also the polarization states
  and wavefront coherence can also be modified.} These effects can be
grouped together into a net complex {\em gain} of the system for each
beam, $\tilde{\mathcal{G}_i}$ -- the amplitude of
$\tilde{\mathcal{G}_i}$ encodes the net amplification or attenuation
of the field strength and the phase term corresponds to combination of
time delays in the system and effects from amplifiers in the signal
chain.  Thus, the measured electric field $\tilde{E}^\prime$ can be
written as a product of the original field $\tilde{E}$ times this
complex gain:
\begin{equation}
  \tilde{E}^\prime = \tilde{\mathcal{G}} \tilde{E} \label{monnier_eqn_1}
\end{equation}

Since the observed complex visibility $\tilde{\mathcal{V}}_{12}$ for a
baseline between telescope 1 to telescope 2 is related to the product
$\tilde{E}_1 \tilde{E}_2^\ast$, then
\begin{eqnarray}
  \tilde{\mathcal{V}}_{12}^\prime & \propto &  \tilde{E}_1^\prime \tilde{E}_2^{\prime\ast} \\
  &\propto & 
\tilde{\mathcal{G}}_1 \tilde{E}_1 \cdot 
                                   \tilde{\mathcal{G}}_2^\ast \tilde{E}_2^\ast \\
      & \propto & \tilde{\mathcal{G}}_1 \tilde{\mathcal{G}}_2^\ast 
                     \tilde{\mathcal{V}}_{12} \label{monnier_eqn_2}
\end{eqnarray}
Thus the measured complex visibility
$\tilde{\mathcal{V}}_{12}^\prime$ is closely related to the true
$\tilde{\mathcal{V}}_{12}$, differing only by complex factor
$\tilde{\mathcal{G}}_1 \tilde{\mathcal{G}}_2^\ast $.  By observing a
calibrator with known structure, this gain factor can be measured,
even if the calibrator is not a point source for the interferometer.
For  a radio array, the gain factors are mainly associated with the
individual telescope collectors and not the baseline, and so the same
gain factors appear in many baselines. This redundancy has led to the
development of additional off-line procedures to "self-calibrate"
radio imaging data  using ``closure amplitude'' techniques (see \S\ref{closurephase}).

\begin{figure}[hbt]
\begin{center}
\includegraphics[angle=90,width=6in]{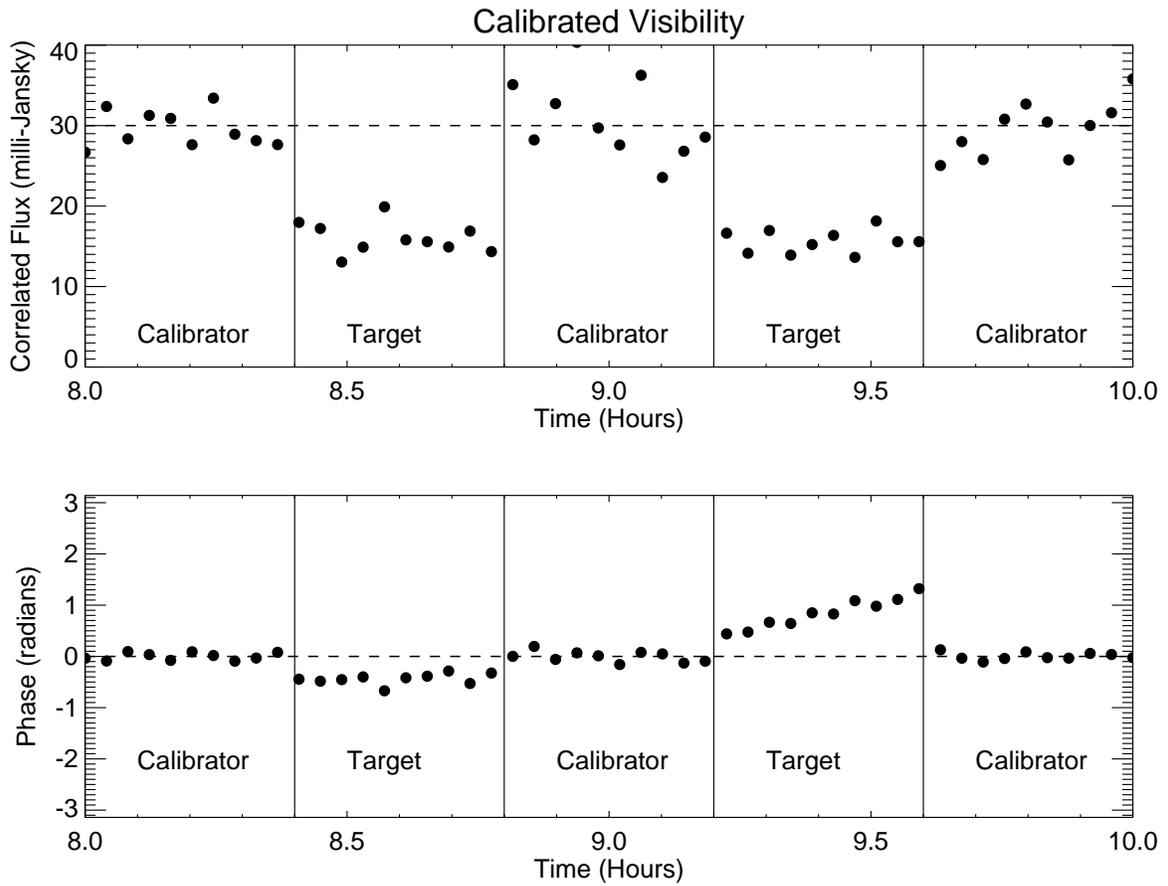}
\figcaption{\footnotesize Calibrator data following estimate of time-varying system visibilites.
\label{calsrccal_b}}

\end{center}
\end{figure}

Once  the system drifts have been estimated  by measurements of the
calibrator,  this correction can be applied to the whole dataset.
Figure~\ref{calsrccal_b} shows the calibrated result, where the
calibrator flux was assumed to be 30~Jy.  In practice, radio phase
calibrators are time-variable in flux and so each dataset typically
includes an ``amplitude calibrator,'' a well-studied object with known
flux as a reference. These calibrated data can now be averaged
and used for further model fitting or synthesis
imaging.  In the example shown here, both the target and calibrator
have reasonable signal-to-noise-ratio. In a more realistic case, the
signal-to-noise of the target will be orders of magnitude worse --
indeed, in one observing block there may be no discernable signal at
all!  The calibrator measurements are used to phase up the array and
allow for very long phase-coherent integrations (averaging in the
complex (u,v) plane).  Unfortunately, this ``blind'' phase referencing
can not generally be used in optical interferometry (see
\S\ref{phasereferencing}) where the short atmospheric coherence time
and the worse turbulence requires active fringe tracking for both
target and calibrator at all times.

Note that actual data will not look quite like this simplified
schematic. First, raw data might have random data glitches or bad
values that need to be flagged. Also, one tends to only observe the
calibrator for a short time, just enough to measure the phase. In fact, the time to
slew between targets can be similar to the length of time spent
integrating on each calibrator. The time
spent on the target during each visit is generally as long as possible given the
atmospheric coherence time which can vary greatly with baseline
length, observing conditions, and wavelength (see \S\ref{turbulence}).  

A common complication is that the calibrator may not be an unresolved
object nor constant in flux.  NRAO maintains a calibrator database
that is used to determine the suitability of each calibrator for
different situations.  As long as the calibrator morphology is known,
the observer can apply a visibility amplitude and phase correction to
account for the known structure.  After this correction, the
calibration procedure is the same.

For visible and infrared interferometry, the procedure is very
similar.  In general, optical interferometers measure a time-averaged
squared-visibility and not visibility amplitude since the
$\mathcal{V}^2$ can be bias-corrected more easily for low
signal-to-noise ratio data when observing with no phase referencing
\citep{colavita1999}.
As discussed earlier, optical interferometers cannot employ phase
referencing between two targets \footnote{Phase referencing is
  possible using ``Dual Star Feeds'' which allows truly simultaneous
  observing of a pair of objects in the same narrow isoplanatic patch
  on the sky.  This capability has been demonstrated on PTI, Keck, and
  VLTI.}  due to the tiny isoplanatic patch and short temporal
coherence times.  Instead of averaging fringe phases, closure phases
(see \S\ref{closurephase}) are formed and averaged over longer time
frames following a similar interpolation of calibration data.
Lastly, calibrators tend to be stars with known or well-estimated
diameters.  For a given baseline, the observed raw calibrator
$\mathcal{V}^2$ are boosted to account for partially resolving them
during the observation before the system visibility is estimated.

When carrying out spectral line and/or polarization measurements,
additional calibrations are required.  As for single telescope
observations, one must observe a source with known spectrum and/or
polarization signature in order to correct for system gains.  These
procedures add steps to the data reduction but are straightforward.

A diversity of packages and data analysis environments are in use for
data reduction of interferometer observations. In the radio and mm-wave
regime, the most popular packages are AIPS and Miriad.  In addition,
the CASA package will be used for ALMA and supports many EVLA
operations now too.  In the visible and infrared, the data reduction
packages are usually closely linked to the instrument and are provided
by the instrument builders.  In most cases there are data analysis
``cookbooks'' that provide step-by-step examples of how to carry out
all steps in the data reduction. Few instruments have complete
pipelines that require no user input, although improved scripting is a
high priority for future development.  A number of summer schools are
offered that train new users of interferometer facilities in the
details of observation planning and data reduction.

The data products from this stage are calibrated complex visibilities
and/or closure phases.  No astronomical interpretation has occurred
yet.  The {\em de facto} standard data format for radio data is
UVFITS.  Unfortunately this format is not strictly defined but rather
represents the data supported by the NRAO AIPS package and importable
into CASA (which uses a new format called the {\em Measurement Set}).
Recently the VLBA community fully documented and registered the FITS
Interferometry Data Interchange (FITS-IDI) format (see {\em ``FITS-IDI
  definition document'' AIPS Memo 114}).  The optical interferometry
community saw the problems of radio in having a poorly defined
standard and, through IAU-sanctioned activites, crafted a common
FITS-based data standard called OIFITS whose specifications were
published by \citet{pauls2005}.  This standard is in wide use by most
optical interferometers in the world today.

   \subsection{Model-fitting for poor (u,v) coverage}
The ultimate goal of most interferometric observations is to have
sufficient data quality and (u,v) coverage to make a synthesis
image. With high image fidelity, an astronomer can interact with the
image just as one would had it come from a standard telescope imager.   

Still, cases are  plentiful where this ideal situation is not
achievable and one will fit a model to the visibility
data directly. There are two classes of models: {\em geometric } models and
{\em physical } models.  {\em Geometric models} are simple shapes that
describe the emission but without any physics involved.  Common examples
include Gaussians, uniform disks, binary system of 2 uniform disks,
etc.  {\em Physical models} start with a physical picture including
densities, opacities, and sources of energy.  Typically a radiative
transfer calculation is used to create a synthetic image which can be
Fourier-transformed (following Equation \ref{eq:vcz2}) to allow fitting to complex visibilities.
Geometric models are useful for very simple cases when an object is
marginally resolved or when physical models are not available (or not
believable!).  Physical models are required to connect observations
with ``real'' quantities like densities and temperatures, although
size scales can be extracted with either kind of model.

In radio and mm-wave, model fitting is now relatively
rare\footnote{Historically, model-fitting was the way data was
  handled in order to discern source structure in the early days of
  radio interferometry. A classic example is the model of Cygnus A,
  which Jennison and Das Gupta fitted to long-baseline
  intensity-interferometry data at 2.4-m wavelength in 1953 \citep{JennisonGupta1953}. \citet[][page 353 et. seq]{Sullivan09} gives more details of
  this fascinating story.}  since high-fidelity imaging is often
achievable.  However in the optical, model fitting is still the most
common tool for interpreting interferometry data.  In many cases, a
simple uniform disk or Gaussian is adequate to express the
characteristic size scale of an object.  By directly fitting to
visibility amplitudes,  the data can be optimally used  and 
proper error analysis can be performed.  The fitting formulae for the two most common
functions can be expressed in closed form as a function of baseline
$B$ and wavelength $\lambda$:
\begin{eqnarray}
 |\mathcal{V}| & = & 2 \frac{J_1 (\pi B \Theta_{\rm diameter} / \lambda)}{\pi B \Theta_{\rm diameter} / \lambda}
    \qquad {\rm case: Uniform~Disk } \\
 |\mathcal{V}| & = & e^{-\frac{\pi^2}{4 \ln{2}} (\Theta_{\rm FWHM} B / \lambda)^2}  \qquad {\rm case: Gaussian }
\end{eqnarray}

\begin{figure}[hbt]
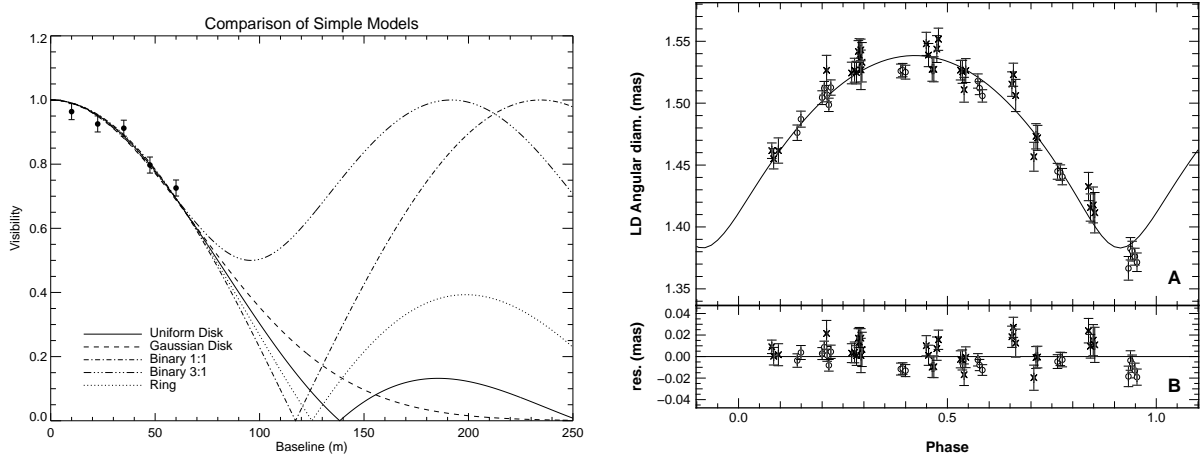

\begin{center}
\includegraphics[angle=90,width=3in]{fig_vismodel.epsi}
\hphantom{...}
\includegraphics[angle=0,width=3in]{fig_merand2005.epsi}
\figcaption{\footnotesize (a) This figure shows visibility data and
  model fits for various brightness distributions, including uniform
  disk, Gaussian disk, binary system and ring.  Notice that all the
  models fit the data nearly identically for visibility about 50\%,
  corresponding here to baselines shorter than 50~m.  One {\em
    requires} long baselines (or exceedingly high signal-to-noise
  ratios) to distinguish between any of these various models,
  illustrating the importance of {\em a priori} knowledge of the
  nature of the target before deciding on appropriate geometrical
  model description.  This figure essentially illustrates in Fourier
  Space the common sense idea that blurry objects all look the same
  unless the source structure is larger than the blur -- here, longer
  baselines essentially reduce the blur of diffraction. (b) The right
  panel shows precision limb-darkened (LD) diameter measurements
  of the classical Cepheid
  $\delta$~Cep at K band through a full pulsational cycle.  These
  tiny ($\pm$4\%) changes were easily tracked by the FLUOR combiner at
  CHARA using $\sim$250-300m baselines \citep[reprinted with
  permission,][]{merand2005}.
\label{fitting_examples}}
\end{center}
\end{figure}

Figure~\ref{fitting_examples}a illustrates both the model-fitting
process and the importance of choosing the most physically-plausible
function.  Here some simulated visibility data spanning baselines from
0 to 60~m are plotted along with curves for 5~common brightness
distributions -- a uniform disk, a Gaussian disk, two binary models
and a ring model. For the case of marginally resolved objects, only
the {\em characteristic scale} of a given model can be constrained and there
is no way to  distinguish {\em between models} without longer baseline information \citep[for more elaborate
discussion, see][]{lachaume2003}.  Note how all the curves fit the
data equally well at short baselines and high visibilities, but that
the {\em interpretation} of each curve is quite different.  Without
longer baselines that can clearly distinguish between these models,
the observer must rely on theoretical expectations to guide model
choice. For example a normal G star should closely resemble a uniform
disk in the near-infrared while disk emission from a young stellar
object in the sub-mm might be more Gaussian.  Despite the
uncertainties when fitting to marginally resolved targets, fitting
interferometric data allows very precise determinations of model
parameters as can be seen in Figure~\ref{fitting_examples}b where the
CHARA Array was used to monitor the variations in diameter of the
Cepheid variable $\delta$~Cep.

\begin{figure}[hbt]
\begin{center}
\includegraphics[angle=0,width=5in]{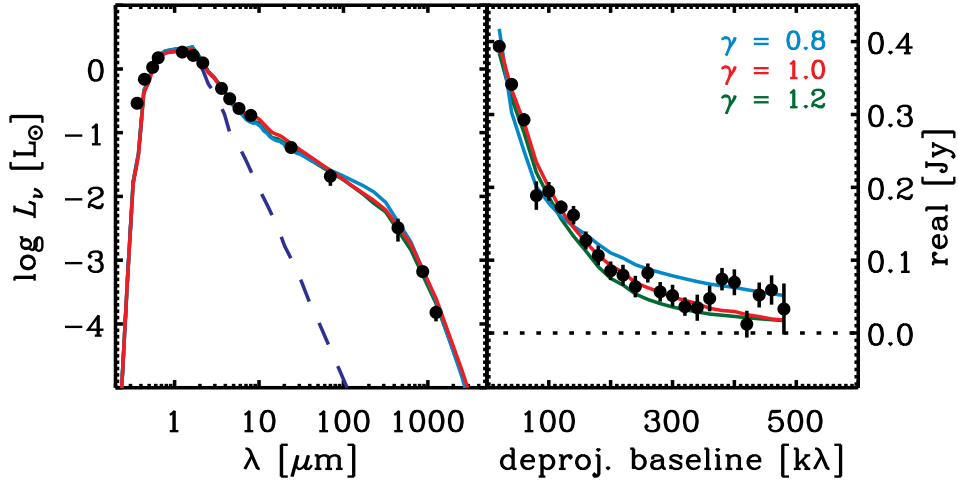}
\figcaption{\footnotesize This figure illustrates how one can fit spectral energy distributions
and visibility simultaneously using simple physical models, as opposed to using
purely descriptive or geometric models.  Models for protoplanetary disk emission
with various radial mass density profiles (left panel) are 
represented by different colored lines.  The visibility data (right panel) were
obtained using the Submillimeter Array at 0.87~mm \citep[reprinted with
permission,][]{andrews2009}.
\label{fig_andrews}}
\end{center}
\end{figure}

A recently-published example of model-fitting at longer wavelengths is shown in
Figure~\ref{fig_andrews}.  Here,  a semi-analytic physical model (in
this case of a circumstellar disk) was used to simultaneously fit
the spectral energy distribution along with the visibility data.  When
realistic physical models are available, multi-wavelength constraints
can make a dramatic improvement to the power of high angular resolution
data and should be included whenever possible.

   \section{Synthesis imaging}
\label{imaging}
Interferometer data in their raw form are not easy to
visualize. Fortunately, as discussed in \S\ref{vcz}, the measured
complex visibilities can be transformed into an equivalent brightness
distribution on the sky -- {\em an image}.  This procedure is called
``aperture synthesis imaging'' or more generally ``synthesis
imaging.'' In this section, the critical data analysis steps are
described for creating an image in both the ideal case as well as more
challenging scenarios when faced with poor (u,v) coverage and phase
instability.

\subsection{Ideal case}
Under ideal conditions the astronomer will have collected
interferometer data with a large number of telescopes including some
Earth rotation to fill-in gaps in (u,v) coverage (see
\S\ref{uvcoverage}).  In addition, each datum will consist of a fully
calibrated complex visibility -- both amplitude and phase information.
Modern radio arrays such as the VLA and ALMA produce data
of this quality when proper phase-referencing procedures (see
\S\ref{phasereferencing}) are employed.

\begin{figure}[hbt]
\begin{center}
\includegraphics[angle=90,width=6in]{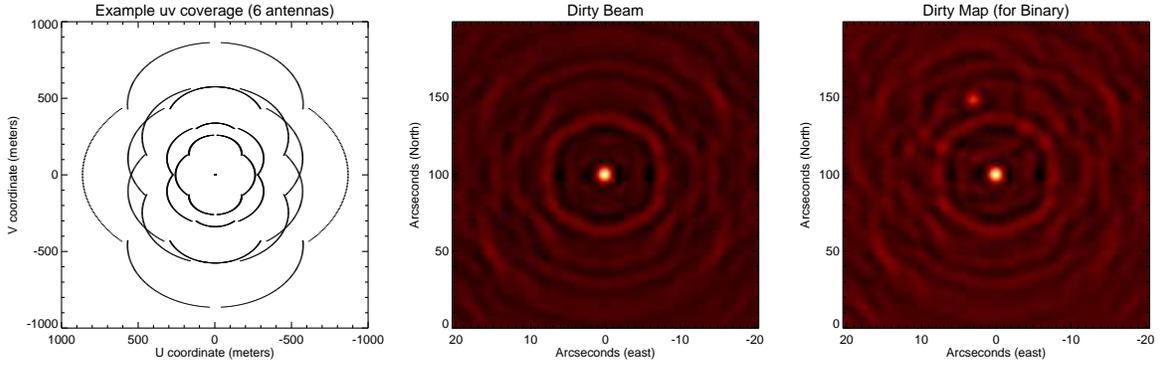}
\figcaption{\footnotesize The basic elements of the CLEAN algorithm consists of (a) the (u,v) coverage of the observations, (b) the
Dirty beam made from the Fourier Transform of the gridded (u,v) support, and (c) the Dirty Map made from Fourier Transform of the observed
(u,v) plane.
\label{fig_imaging1}}
\end{center}
\end{figure}

Figure~\ref{fig_imaging1} depicts the Fourier coverage for a 6
telescope interferometer along with the resulting image from a
direct Fourier transform of this coverage for a perfect point source
(by construction, the image will be purely real).  In this
procedure, the values in the (u,v) grid are set to unity where data
exists and to zero where no data exists.  This resulting image shows
artifacts because of the missing (u,v) data that were zeroed out.
These artifacts are often called ``sidelobes'' and show both positive
and negative excursions -- note that negative flux density is usually
a strong sign of a sidelobe since negative values in an image are
typically unphysical (except for special applications like absorption
line studies or polarization stokes mapping).  Note that the central
core of the image represents the diffraction-limited angular
resolution of this observation.  Often in a practical situation, the
central core will be elongated because (u,v) coverage is not perfectly
symmetric, with longer baselines in some directions than others.  The
last panel in this figure shows the resulting image for a binary star
with 2:1 flux ratio.  Notice that the image contains two sources,
however both sources show the same pattern of artifacts as the simple
point source.  Indeed, this supports the previous assertion that missing (u,v)
coverage is the main origin of this pattern and suggests that the image quality can be
improved by  correcting for the sidelobe effect.

To proceed, the well-known Convolution Theorem must be introduced: {\em
  Multiplication in the ``(u,v) space'' is equivalent to a convolution
  in ``image space.''}  Mathematically, this as can be expressed:

\begin{eqnarray}
   {\rm FT} ( \tilde{\mathcal{V}}(u,v) \cdot M(u,v) ) & = &  ({\rm FT}~\tilde{\mathcal{V}}) \otimes ({\rm FT}~M) \\
   \tilde{\mathcal{V}}(u,v) \cdot M(u,v) & \Leftrightarrow & I(x,y) \otimes B(x,y)
\end{eqnarray}
where $\tilde{\mathcal{V}}$ is the full underlying complex visibility
that could in principle be measured by the interferometer, $M$ is the
(u,v) mask that encodes whether data exists (1) or is missing (0)\footnote{In general, one can consider a full ``Spatial Transfer Function'' which can have weights between 0 and 1.  Here, 
consider just a simple binary mask in the (u,v) plane for simplicity.}, FT and
$\Leftrightarrow$ denote a Fourier Transform, $I={\rm
  FT}~\tilde{\mathcal{V}}$ is the true image distribution , and
$B={\rm FT}~M$ is called the ``Convolving Beam.''

Application of the convolution theorem permits
an elegant reformulation of the  imaging problem into a ``deconvolution'' problem,
where the convolving beam is a complicated function but derived
directly from $M$, the observed (u,v) coverage.  Since $M$ is exactly
known, the convolving beam $B$ is  known as well.  Note that
this deconvolution problem contrasts sharply with the deconvolution
problem in adaptive optics imaging where the point source function
(PSF) varies in time and is never precisely known.

\subsubsection{CLEAN Algorithm}
One of the earliest methods developed for deconvolution was the CLEAN
algorithm \citep{hogbom1974}, which is still widely used in radio
interferometry.  In CLEAN, the Fourier transform of the gridded
complex visibility data is called the ``dirty image'' (or sometimes
``dirty map'') and the Fourier transform of the (u,v) plane mask is
called the ``dirty beam.''  One first needs to deconvolve the dirty
image with the dirty beam.  To do this, the true image $I$ is
interatively constructed by locating the peak in the dirty image and
subtracting from this a scaled version of the dirty beam centered at
this location.  Here, the scaling of the dirty beam is typically tuned
to removed a certain fraction of intensity from the peak, often 5\%.
One keeps track of how much one removes by collecting the ``CLEAN''
components in a list.

Consider this example. The dirty image has peak of 1.0 Jy at pixel
location (3,10).  One creates a scaled dirty beam with a peak of
0.05~Jy, shifts the peak to the position (3,10), and subtracts this
scaled dirty beam from the dirty image. This CLEAN component is
collected and labeled by location (3,10) and flux contribution (0.05
Jy).  Continue this procedure, flux is removed from dirty image and
CLEAN components are collected.  This procedure is halted when the residual
dirty image contains only noise. Since the intensity in the true
image is expected to be positive-definite, one common criterion for halting the
CLEAN cycles is when the largest negative value in the image is
comparable to the largest positive value in the image, thereby
avoiding any CLEAN components having negative flux.

\begin{figure}[hbt]
\begin{center}
\includegraphics[angle=90,width=6in]{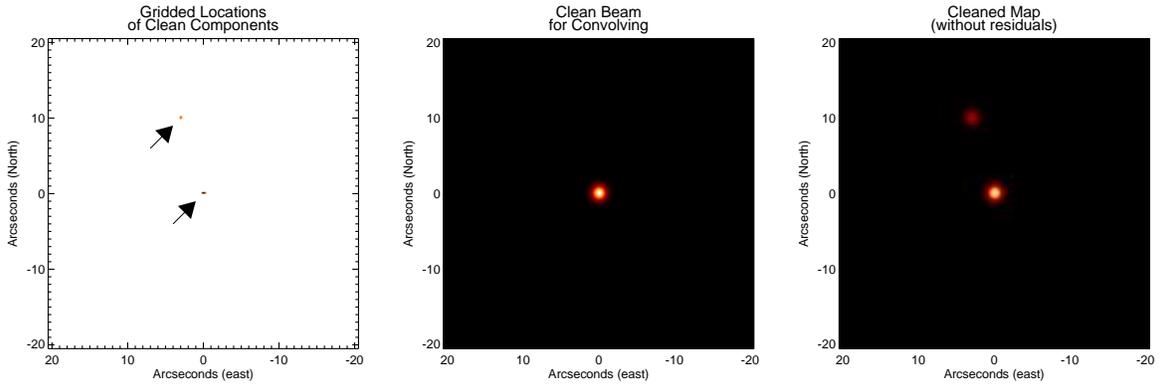}
\figcaption{\footnotesize The basic ``restoration'' steps of the CLEAN algorithm
 consist of (a) collecting the locations of the CLEAN components, (b) 
creating a CLEAN beam which is a Gaussian beam with same angular resolution
of the central portion of the Dirty Beam, and (c)
convolving the CLEAN components with the CLEAN beam.  Here, the final step
of adding back the residuals of the dirty map has been left off.
The arrows in panel (a) mark the location of the two clusters of clean
components near the locations of the two components of the 
binary.
\label{fig_imaging2}}
\end{center}
\end{figure}

In principle, this collection of delta function point sources (all the
CLEAN components) {\em is} the best estimate of the image
distribution. However, an image of point sources is not visually
appealing and a common procedure is to convolve the point source
distribution with a ``CLEAN'' beam, a perfect 2-D Gaussian with a core
that matches the FWHM of the dirty beam.  Commonly, a filled ellipse will be
included in the corner of a CLEANed image showing the 2D FWHM of the
restoring beam.  Lastly, one adds back the residual image from the
dirty map (which should contain only noise) so that the noise level is
apparent in the final image and remnant uncorrected sidelobe
artifacts, if present, can be readily identified. These steps are
illustrated in Figure~\ref{fig_imaging2} where  the
CLEAN procedure has been applied to the examples shown in Figure~\ref{fig_imaging1}.

The steps of the CLEAN algorithm are summarized below.

\begin{enumerate}
\item Create dirty map and beam.
\item Find peak of dirty map.
\item Subtract scaled version of dirty beam from dirty map, removing a small percentage (e.g., $\sim$5\%) of peak intensity.  Collect CLEAN components.
\item Repeat last step until negative residuals are comparable to positive residuals.
\item Convolve CLEAN components with CLEAN beam.
\item Add back the residuals.
\end{enumerate}

While it is beyond the scope to address the weaknesses of the CLEAN
algorithm in detail, a few issues are worth mentioning  in passing.  Typical CLEAN
algorithms do not naturally deal with errors unless all the visibility
data are of similar quality. The case when each visibility point is
weighted equally is called {\em natural weighting} (which gives best
SNR for detecting faint objects) and the case when each portion of the
(u,v) plane is equally weighted, so called {\em uniform weighting}
(which gives somewhat higher angular resolution but at some loss of
sensitivity).  \citet{briggs1995} introduced a ``ROBUST'' parameter
than can naturally span these two extremes.  Other problems with CLEAN
include difficulty reconstructing low surface brightness regions,
large scale emission and the fact that the final reconstructed image
is of a degraded resolution because the convolving PSF actually
suppresses the longest baseline visibilities from the final image.
Finally, if the imaging step makes use of the Fast Fourier Transform,
several additional artifacts can appear in the image as a consequence
of the necessity to grid the input $u,v$ data.

\subsubsection{Maximum Entropy Method (MEM)}

Another common method for reconstructing images from interferometric
data is called the {\em Maximum Entropy Method}
\citep{gull1983,skilling1984}.  This approach asks the question: ``How
to choose which reconstructed image is best, considering that there
are an infinite number of images that can fit the interferometer
dataset within the statistical uncertainties?''  The simple answer
here is that the ``best'' image is the one that both fits the data
acceptably and also maximizes the {\em Entropy} $S$ defined as:

\begin{equation}
S = -\sum_i f_i \ln \frac{f_i}{m_i} 
\end{equation}
\noindent where $f_i$ is the (positive-definite) fraction of flux in
pixel $i$, and $m_i$ is called the {\em image prior} which can
encapsulate prior knowledge of the flux distribution (e.g., known from
physical considerations or lower resolution observations).
\citet{narayan1986} describe general properties of this algorithm in a
lucid review article and motivates the above methodology using {\em
  Bayes' Theorem}, the cornerstone of so-called {\em Bayesian
  Statistics}.

Entropy is an interesting statistic since it quantifies the amount of
complexity in a distribution.  It is often stated that MEM tries to find the
``smoothest image'' consistent with the data. This indeed is a highly
desirable feature since any structures in the reconstructed image
should be based on the data itself and not artifacts from the
reconstruction process.  However, MEM does not actually select the
smoothest image but rather one with ``most equal'' uniform set of
values -- MEM does not explicitly take into account spatial structure
but only depends on the distribution of pixel values.  Indeed with more
study, the Maximum Entropy functional has been found  not to be that
special, except for its privileged appearance in physics. From a broader perspective, maximum entropy can be considered one member of a class of regularizers that
allow the inverse problem to be well-defined, and MEM is not necessarily
the best nor even suitable for some imaging problems (e.g., other regularizers include total variation or maximum likelihood).

MEM performs better for reconstructing smooth large scale emission
than CLEAN, although MEM is much more computationally demanding.  MEM
can naturally deal with heterogeneous data with varying errors since
the data is essentially fitted using a $\chi^2$-like statistic. This
involves only a ``forward'' transform from image space to (u,v) space, thus
avoiding all the issues of zeros in the (u,v) grid and the need for
deconvolution.  In addition, MEM images possess {\em super-resolution}
beyond the traditional $\sim\frac{\lambda}{D}$ diffraction limit since
smoothness is introduced in the process only indirectly through the
Entropy statistic.  For instance, if the object is a point-like
object, then the FWHM of the reconstructed MEM image depends on the
signal-to-noise of the data, not just the length of the longest
baseline.  Super-resolution is viewed with some skepticism by
practioners of CLEAN because structures beyond the formal diffraction
limit may be artifacts of the Entropy functional.  MEM has been
implemented in many radio interferometry data processing environments,
such as AIPS for VLA/VLBA, Caltech VLBI package \citep{sivia1987}, and
for optical interferometry \citep[e.g., BSMEM,][]{buscher1994}.

\subsection{Non-ideal cases for imaging}
Most modern interferometric arrays in the radio (VLA) and millimeter
(ALMA, CARMA) have a sufficient number of elements for good (u,v) 
coverage and also
employ rapid phase referencing for absolute phase calibration.  This
allows either CLEAN and MEM methods to make imaging possible, although
mosaicing large fields can still pose a computational challenge.

In optical interferometry and perhaps also at some sub-mm wavelengths,
atmospheric turbulence changes the fringes phases so quickly and over
such small angular scales that phase referencing is not practical.  As
discussed earlier, modeling can still be done for the visibility
amplitudes but the turbulence scrambles the phase information beyond
utility. Fortunately, there is a clever method to recover some of the
lost phase information and this is discussed next.

\subsubsection{Closure Phase, the Bispectrum, and Closure Amplitudes}
\label{closurephase}
As discussed earlier, phase referencing is used to correct for
drifting phases.  So what can be done when phase referencing is not
possible?  Without valid phase information accompanying the visibility
amplitude measurements, one cannot carry out the inverse Fourier
Transform that lies at the core of synthesis imaging and the
CLEAN algorithm specifically.  While in some cases the fringe phases do
not carry much information (e.g., for symmetrical objects), in
general phases carry most of the information for complex scenes.

Early in the history of radio interferometry a clever idea, now
referred to as ``closure phase,'' was discovered to recover some level
of phase information when observing with three telescopes
\citep{jennison1958}.  The method was introduced to partly circumvent
the combination of poor receiver stability and variable multi-path
propagation in early radio-linked long-baseline ($ \gtrsim 2$ km)
interferometer systems at Jodrell Bank. The term ``closure phase''
itself appeared later on, in the paper by \citet{rogers1974}
describing an application at centimeter radio wavelengths using very
stable and accurate, but independent, reference oscillators at the
three stations in a so-called very long baseline interferometer (VLBI)
array.  Closure phase was critical for VLBI work in the 1980s although
it became less necessary as phase referencing became feasible.
Application at optical wavelengths was first mentioned by
\citet{rogstad1968}, but carried out only much later in the optical
range through aperture masking experiments
\citep[e.g.,][]{baldwin1986}.  By 2006, nearly all separated-element
optical arrays with 3~or more elements have succeeded in obtaining
closure phase measurements (COAST, NPOI, IOTA, ISI, VLTI, CHARA).  An
optical observer now can expect closure phases to be a crucial
observable for most current instrumentation.

The principle behind the power of closure phases is briefly described
and the interested reader is referred to \citet{monnier2007} for more
detailed information on taking advantage of such phase information in
optical interferometry.

\begin{figure}[t]
\centering
\includegraphics[height=3in,clip]{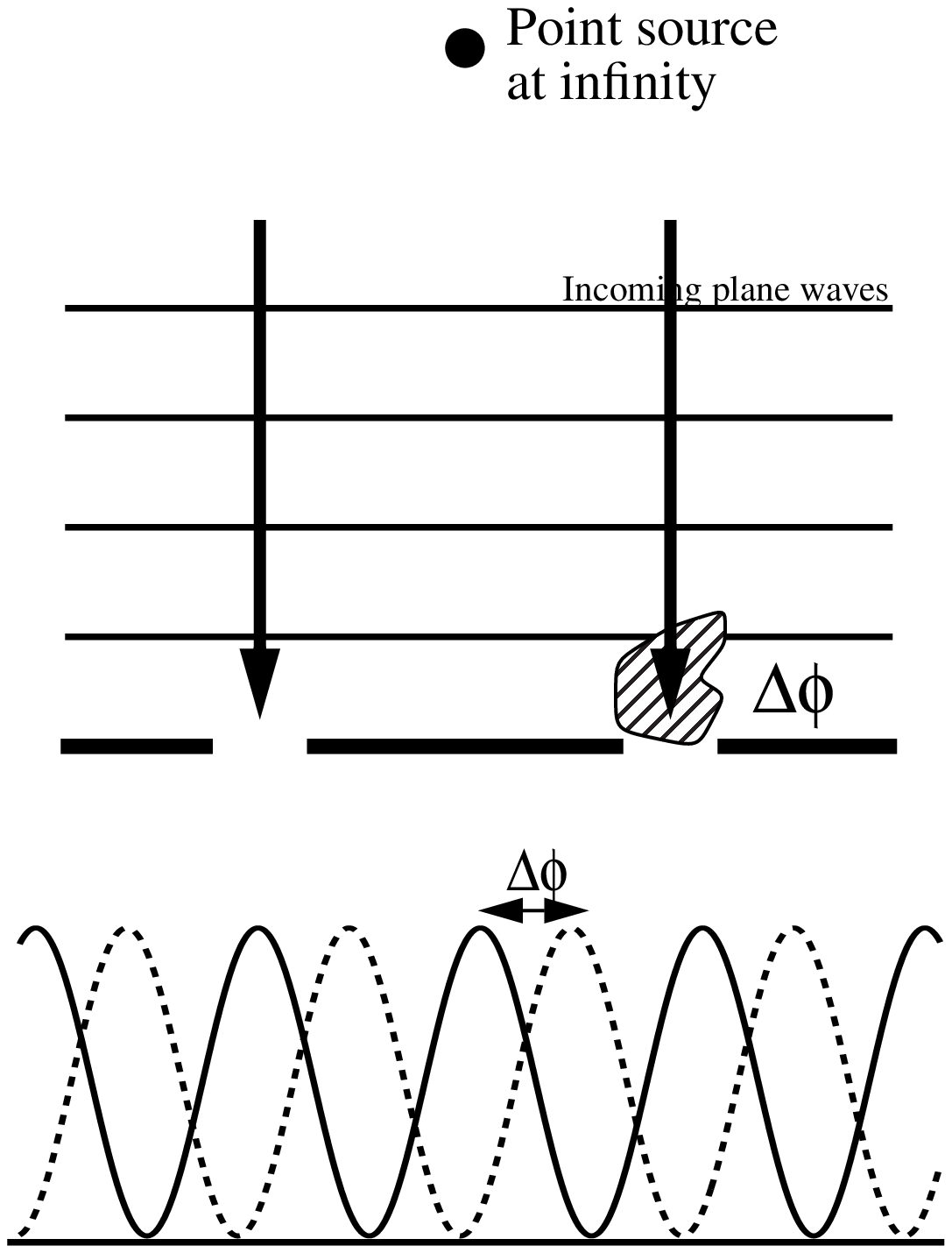}
\hphantom{.....}
\includegraphics[clip,height=3in]{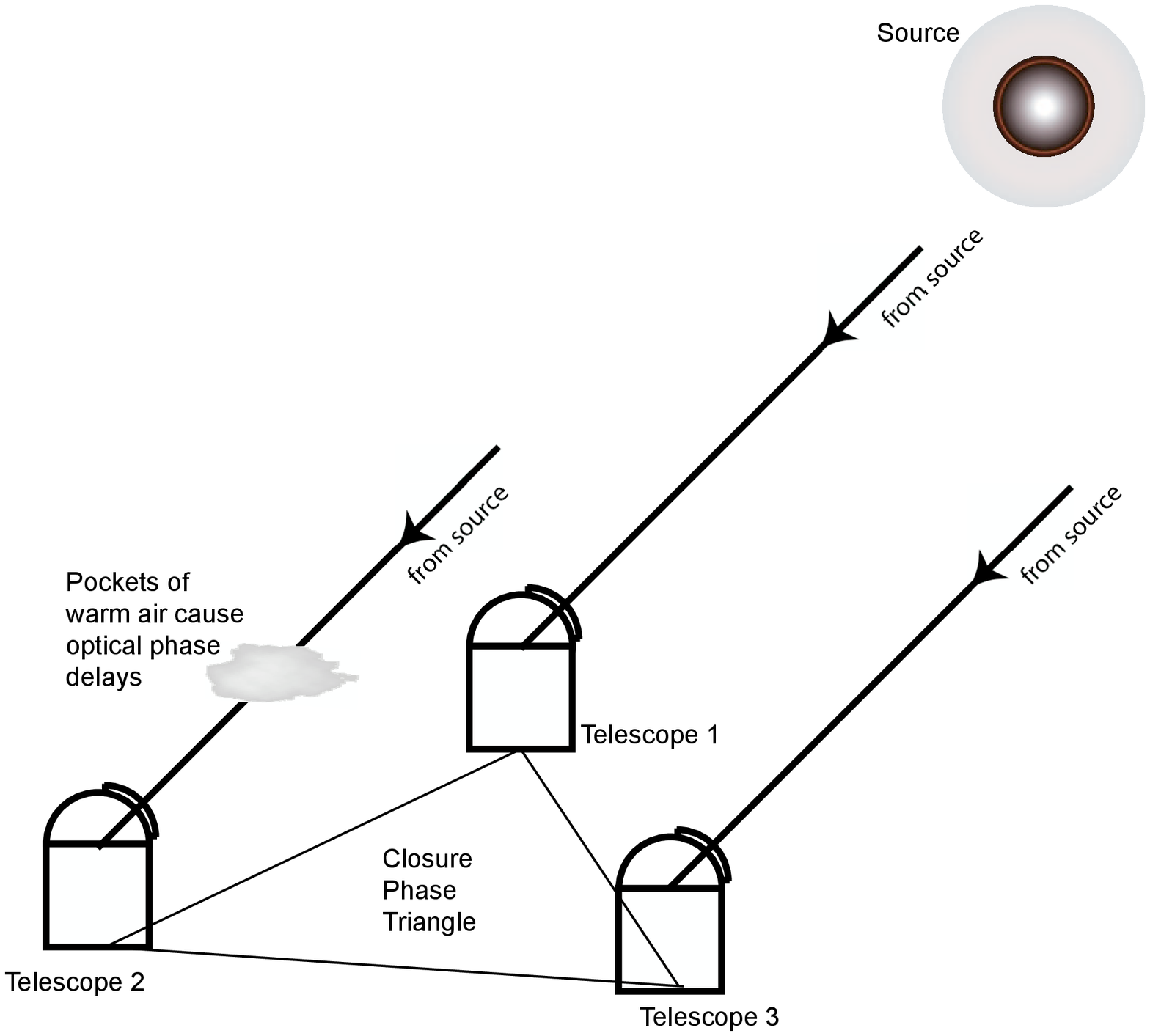}
\caption{(left) Atmospheric turbulence introduces extra path length fluctuations that
induce fringe phase shifts. At optical wavelengths, these phase shifts vary by many radians over short time scales ($<<$1 sec) effectively scrambling the Fourier phase information. 
(right) Phase errors introduced at any telescope causes equal but opposite phase
shifts in ajoining baselines, canceling out in the {\em closure phase}
\citep[see also][]{readhead1988,ionic3_2006}.
\citep[figures reprinted from][]{monnier2003}
\label{cphasefig}
}
\end{figure}

Consider Figure~\ref{cphasefig}a in which a time delay is introduced
above one slit in a Young's interferometer.  This time delay
introduces a phase shift for the detected fringe and the magnitude of
the phase shift is independent of the baseline length.  For the case
of 3~telescopes (see Figure~\ref{cphasefig}b),  a delay above
one telescope will introduce phase shifts in {\em two fringes}.  For
instance, a delay above telescope 2 will show up as an equal phase
shift for baseline 1-2 and baseline 2-3, but with {\em opposite}
signs.  Hence, the sum of three fringe phases, between 1-2, 2-3, and
3-1, will be insensitive to the phase delay above telescope 2.  This
argument holds for arbitrary phase delays above any of the three
telescopes.  In general, the sum of three phases around a closed
triangle of baselines, the {\em closure phase}, is a good
interferometric observable; that is, it is independent of
telescope-specific phase shifts induced by the atmosphere or optics.

The closure phase $\Phi_{ijk}$ can thus be written in terms of the three
telescopes $i$,$j$,$k$ in the triangle:

\begin{equation}
\Phi_{ijk} = \phi_{ij} + \phi_{jk} + \phi_{ki}
\end{equation}
\noindent where $\phi_{ij}$ represents the measured Fourier phase for the
baseline connecting telescopes $i$,$j$.  Alternatively, the closure
phase can be written in terms of the ($u_0$,$v_0$,$u_1$,$v_1$) in the
Fourier (hyper-)plane where ($u_0$,$v_0$) represents the (u,v)
coverage for baseline $i,j$ in the triangle, ($u_1$,$v_1$)
represents the (u,v) coverage for baseline $j,k$ in the triangle, and
the last leg of the triangle can be calculated from the others since
the sum of the 3 baselines must equal zero to be a ``closure
triangle.''  See definition and explanation put forward in
documentation of the OI-FITS data format \citep{pauls2005}.

Another method to derive the invariance of the closure phase to
telescope-specific phase shifts is through the {\em bispectrum}.
The bispectrum $\tilde{B}_{ijk}=  \tilde{\mathcal V}_{ij}
\tilde{\mathcal V}_{jk}\tilde{\mathcal V}_{ki}$
is formed through triple products of the complex
visibilities around a closed triangle, where $ijk$ specifies the
three telescopes.
Using Eq. \ref{monnier_eqn_2} and using the concept of telescope-specific complex gains $G_i$,
it can be seen how the telescope-specific errors affect the measured bispectrum:
\begin{eqnarray}
\tilde{B}_{ijk} & = &  \tilde{\mathcal V}^{\meas}_{ij} \,
\tilde{\mathcal V}^{\meas}_{jk}\,
\tilde{\mathcal V}^{\meas}_{ki} \\
& = & |G_i| |G_j| e^{i (\Phi^G_i -\Phi^G_j)}
\tilde{\mathcal V}^{\true}_{ij} \cdot
|G_j| |G_k| e^{i (\Phi^G_j -\Phi^G_k)}
\tilde{\mathcal V}^{\true}_{jk} \cdot
|G_k| |G_i| e^{i (\Phi^G_k -\Phi^G_i)}
\tilde{\mathcal V}^{\true}_{ki} \\
& = &
|G_i|^2 |G_j|^2 |G_k|^2
\tilde{\mathcal V}^{\true}_{ij} \cdot
\tilde{\mathcal V}^{\true}_{jk} \cdot
\tilde{\mathcal V}^{\true}_{ki}
\end{eqnarray}

From the above derivation, the bispectrum is a complex quantity whose
phase is identical to the closure phase, while the individual
telescope gains affect only the bispectrum amplitude.  The use of the
bispectrum for reconstructing diffraction-limited images from speckle
data was developed independently \citep{weigelt1977} of the closure
phase techniques, and the connection between the approaches elucidated
only later \citep{roddier1986,cornwell1987}.

A 3-telescope array with its one triangle can provide a single closure
phase measurement, a paltry substitute for the 3 Fourier phases
available using phase referencing.  However, as one increases the
number of elements in the array from 3 telescopes to 7 telescopes, the
number of independent closure phases increases dramatically to 15,
about 70\% of the total 21 Fourier phases available.  An array the
size of the VLA with 27 antennae is capturing 93\% of the phase
information. Indeed, imaging of bright objects does not require phase referencing
and the VLA can make high quality imaging through closure phases
alone. Note that imaging using closure phases alone retains no
absolute astrometry information; astrometry requires
phase-referencing.

A related quantity useful in radio is the {\em closure amplitude}
(which requires sets of 4~telescopes) and this can be used to compensate for
unstable amplifier gains and varying antenna efficiencies \citep[e.g.,][]{readhead1980}.
Closure amplitudes are not practical for current optical
interferometers partially because most fringe amplitude variations are
not caused by telescope-specific gain changes but rather by changing
coherence (e.g., due to changing atmosphere).

Closure phases (and closure amplitudes) can be introduced into the
imaging process in a variety of ways.  For the CLEAN algorithm, the
closure phases can not be directly used because CLEAN requires
estimates of the actual Fourier phases in order to carry out a Fourier
transform.  A clever iterative scheme known as {\em self-calibration}
was described by \citet{readhead1978} and \citet{cornwell1981} which alternates
between a CLEANing stage and a self-calibration stage that estimates
Fourier phases from the closure phases and most recent CLEANed image.
As for standard CLEAN itself, self-calibration can not naturally deal with
errors in the closure phases and closure amplitudes, and thus is recognized as not
optimal.  That said, self-calibration is still widely used along with
CLEAN even in the case of phase referencing to dramatically improve
the imaging dynamic range for imaging of and around bright objects.

Closure phases, the bispectrum, and closure amplitudes can be quite
naturally incorporated into ``forward-transform'' image reconstruction schemes such as the
Maximum Entropy Method.  Recall that MEM basically performs a
minimization of a regularizer constrained by some
goodness-of-fit to the observed data. Thus, the bispectral quantities
can be fitted just like all other observables. That said, the
mathematics can be difficult and the program BSMEM \citep{buscher1994}
was one of the first useful software suitable for optical interferometers
that successfully solved this problem in practice.  Currently, the
optical interferometer community have produced several algorithms to
solve this problem, including the Building Block Method \citep{hofmann1993}, MACIM \citep{macim}, MIRA
\citep{thiebaut2008}, WISARD \citep{meimon2008}, and SQUEEZE
\citep{baron2010}.  See \citet{malbet2010} for a description of a
recent blind imaging competition between some of these algorithms.

\subsection{Astrometry}

Astrometry is still a specialized technique within interferometry and
a detailed description is beyond the scope of this chapter.
Typically, the precise separation between two objects on the sky, the
{\em relative astrometry}, is needed to be known for some purpose,
such as a parallax measurement.  If both objects are known
point-like objects with no asymmetric structure, then the
precise knowledge of the baseline geometry can be used, along with detailed measurements
of interferometric fringe phase, to estimate their angular separation.
In general, the astrometric precision $\Delta\Phi$ on the sky is
related to the measured fringe phase SNR as follows:
\begin{equation}
\Delta\Phi \sim \frac{\lambda}{\rm Baseline} \cdot \frac{1}{\rm SNR}
\end{equation}

Hence, if one measures a fringe with a signal-to-noise of 100 (which
is quite feasible) then one can determine the relative separation of
two sources with a precision 100$\times$ smaller than the fringe
spacing.  Since VLBI and optical interferometers have fringe spacings
of $\sim$1~milliarcseconds, this allows for astrometric precision at
the 10~{\em microarcsecond} level allowing parallax measurements at
many kiloparsecs and also allows us to take a close look around  the black hole at
the center of the Galaxy.  Unfortunately, at this precision level many
systematic effects become critical and knowledge of the absolute
baseline vector between telescopes is crucial and demanding.  The 
interested reader should consult specific instruments and recent results
for more information \cite[][for radio and optical
  respectively]{reid2009,muterspaugh2010a}.

\section{Concluding remarks}

In barely more than 50 years, separated-element interferometry has
come to dominate radio telescope design as the science has demanded
ever-increasing angular resolution. The addition of more elements to
an array has provided improvements both in mapping speed and in
sensitivity. The early problems with instability in the electronics
were solved years ago by improvements in radio and digital
electronics, and it is presently not unusual to routinely achieve
interferometer phase stability of order $\lambda/1000$ over time spans
of many tens of minutes. Sensitivity improvements by further
reductions in receiver temperature are reaching a point of diminishing
returns as the remaining contributions from local spillover, the
Galactic background, and atmospheric losses come to dominate the
equation. Advances in the speed and density of digital solid-state
devices are presently driving the development of increased
sensitivity, with flexible digital signal processors providing wider
bands for continuum observations and high frequency resolution for
precision spectroscopy.  With the commissioning of the billion-dollar
international facility ALMA underway and long-term plans for ambitious
arrays at longer wavelengths (SKA, LOFAR), the future of ``radio''
interferometry is bright and astronomers are eager to take advantage
of the new capabilities, especially vast improvements to sensitivity.

The younger field of optical interferometry is still rapidly
developing with innovative beam combination, fringe tracking methods
and extensions of adaptive optics promising significant improvements
in sensitivity for years to come.  In the long-run however, the
atmosphere poses a fundamental limit to the ultimate astrometric
precision and absolute sensitivity for visible and infrared
interferometry that can only be properly overcome by placing the
telescopes into space or through exploring exotic sites such as Dome
A, Antarctica. Currently however, the emphasis of the majority of the
scientific community favors ever-larger collecting area in the near-
and mid-IR (e.g.\ JWST) as astronomers reach to ever greater distances and
earlier times, rather than higher angular resolution.

 Eventually, the limits of diffraction are likely
to limit the science return of space telescopes, and the traditional
response of building a larger aperture will no longer be
affordable. Filled apertures increase in weight and in cost as a power
$> 1$ of the diameter \citep{bely2003}. For ground-based telescopes,
cost $\sim D^{2.6}$. In space, the growth is slower, $\sim D^{1.6}$,
although the coefficient of proportionality is much
larger\footnote{It's interesting to note that at some large-enough
  diameter, ground-based telescopes are likely to become similar in
  cost to those in space, especially if one considers life-cycle
  costs.}. Over the last two decades the prospects have become
increasingly bleak for building ever larger filled-aperture telescopes
within the anticipated space science budgets. Interferometry permits the connection
of individual collectors of modest size and cost into truly gigantic
space-based constellations with virtually unlimited angular resolution
\citep{allen2007}. Ultimately, the pressure of discovery is likely to make
interferometry in space at optical, UV, and IR wavelengths a
necessity, just as it did at radio wavelengths more than half a
century earlier.

\appendix
\section{Appendix: Current Facilities}
\label{appendix}

Table \ref{table:all_interferometers} provides a comprehensive list of
existing and planned interferometer facilities. These facilities span
the range from ``private'' instruments, currently available only to
their developers, to general purpose instruments. An example of the
latter is the Square Kilometer Array; this is a major next-generation
facility for radio astronomy being planned by an international
consortium, with a Project Office at the Jodrell Bank Observatory in
the UK (http://www.skatelescope.org/). The location for the SKA has
not yet been finalized, but (as of 2011) sites in Australia and South
Africa are currently being discussed. Plans for the SKA have spawned
several "proof-of-concept" or "pathfinder" instruments which are being
designed and built including LOFAR (Europe), MeerKAT (South Africa),
ASKAP, MWA, and SKAMP (Australia), and LWA (USA).

\begin{figure}[t]
\centering
\includegraphics[angle=90,width=6in]{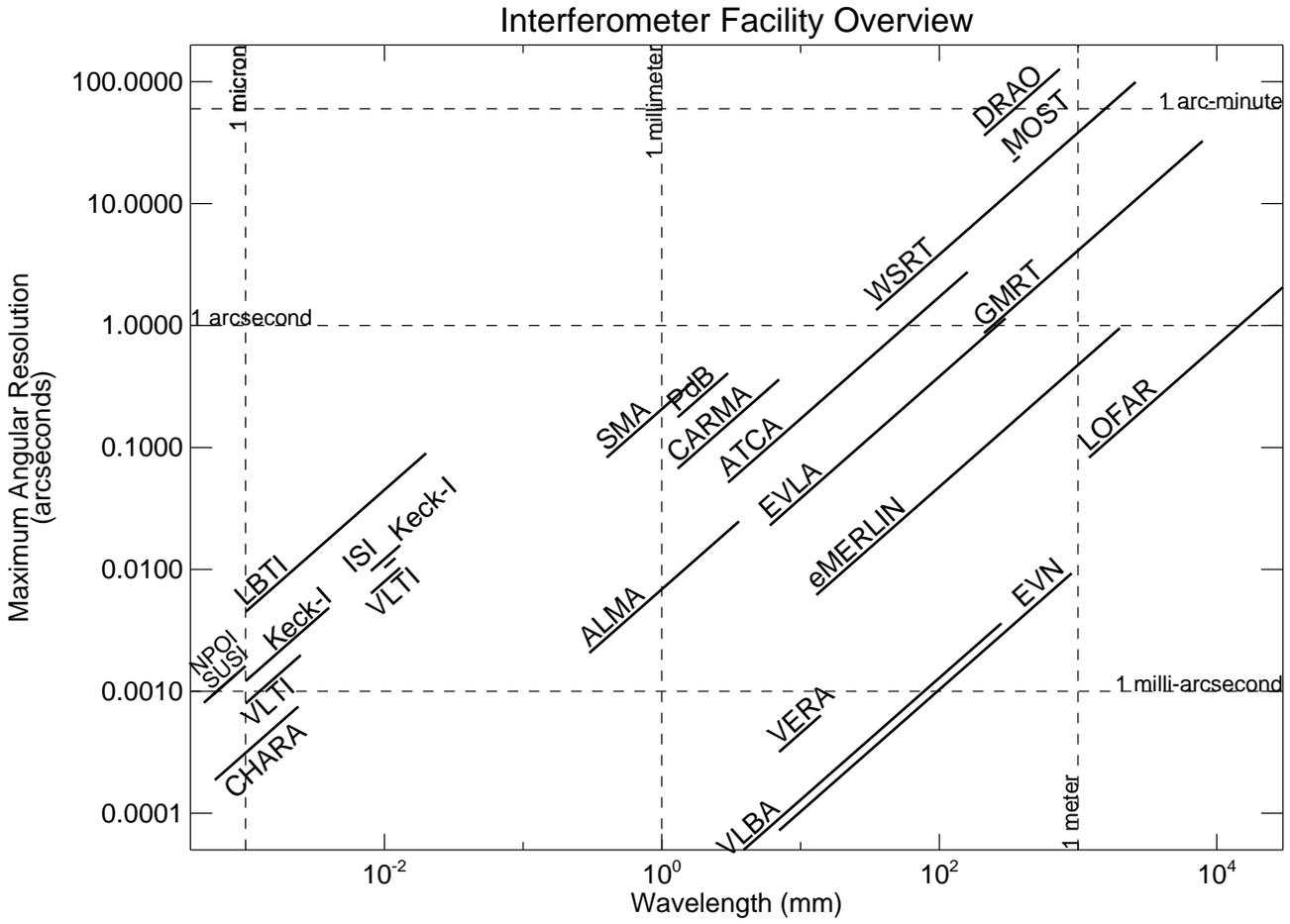}
\caption{Graphical representation of the wavelength coverage and maximum angular resolution available using the radio and optical interferometers of the world. See Table~\ref{table:capabilities} for more information.
\label{fig_capabilities}
}
\end{figure}

Table~\ref{table:capabilities} summarizes the wavelength and angular
resolution of currently operating arrays open to the general
astronomer.  A few of the facilities in this list are still under
construction.  Figure~\ref{fig_capabilities} summarizes the vast
wavelength range and angular resolution available using the radio and
optical interferometers of the world.

\begin{landscape}
\begin{table}
\footnotesize
\caption {Alphabetical list of current and planned interferometer arrays. \label{table:all_interferometers}}
\begin{center}
\begin{tabular}{|l|l|l|l|}
\hline
Acronym & Full Name & Lead Institution(s) & Location \\ \hline
ALMA & Atacama Large Millimeter Array & International/NRAO & Chajnantor, Chile \\
ATA & Allen Telescope Array & SETI Institute & Hat Creek Radio Observatory, CA, USA \\
ATCA & Australia Telescope Compact Array & CSIRO/ATNF & Narrabri, Australia \\
CARMA & Combined Array for Research & Caltech, UCB, UChicago, & Big Pine, California USA \\
 & in Millimeter Astronomy & UIUC, UMD & \\
CHARA & Center for High Angular Resolution Astronomy & Georgia State University &  Mt. Wilson, CA, USA \\
DRAO & Dominion Radio Astrophysical Observatory & Herzberg Institute of Astrophysics & Penticton BC, Canada \\
EVLA & Expanded Very Large Array & NRAO & New Mexico USA \\
EVN & The European VLBI Network & International & Europe, UK, US, S. Africa, China, Russia \\
FASR & Frequency Agile Solar Radiotelescope & National Radio Astronomy Observatory + 6 others & Owens Valley, CA, USA \\
ISI & Infrared Spatial Interferometer & Univ.\ California at Berkeley & Mt. Wilson, CA, USA \\
Keck-I & Keck Interferometer (Keck-I to Keck-II) & NASA-JPL & Mauna Kea, HI, USA \\
LBTI & Large Binocular Telescope Interferometer& LBT Consortium & Mt. Graham, AZ, USA \\
LOFAR & Low Frequency ARray & Netherlands Institute & Europe \\
 & & for Radio Astronomy - ASTRON & \\
LWA & Long Wavelength Array & US Naval Research Laboratory & VLA Site, NM, USA \\
MOST & Molonglo Observatory Synthesis Telescope & School of Physics, Univ.  Sydney & Canberra, Australia \\
MRO & Magdalena Ridge Observatory & Consortium of New Mexico Institutions, & Magdalena Ridge, NM, USA \\
MWA &  Murchison Widefield Array & International/MIT-Haystack & Western Australia \\
NPOI & Navy Prototype Optical Interferometer & Naval Research Laboratory/ &
Flagstaff, AZ, USA \\
 & & U.S. Naval Observatory & \\
OHANA & Optical Hawaiian Array & Consortium (mostly French Institutions), & Mauna Kea, HI, USA \\
 & for Nanoradian Astronomy  & Mauna Kea Observatories, others & \\
PdB & Plateau de Bure interferometer & International/IRAM & Plateau de Bure, FR  \\
SKA & Square Kilometer Array & see Appendix \ref{appendix} & under study \\
SMA & Submillimeter Array & Smithsonian Astrophysical Observatory & Mauna Kea, HI, USA \\
SUSI & Sydney University Stellar Interferometer & Sydney University & Narrabri, Australia \\
VLBA & Very Long Baseline Array & NRAO & Hawaii to St. Croix, US Virgin Islands \\
VLTI-UT & VLT Interferometer (Unit Telescopes) & European Southern Observatory & Paranal, Chile \\
VLTI-AT & VLT Interferometer (Auxiliary Telescopes) & European Southern Observatory & Paranal, Chile \\
VERA & VLBI Exploration of Radio Astrometry & Nat. Astron. Obs. Japan (NAOJ) & Japan \\
VSOP-2 & VLBI Space Observatory Programme-2 & Japan Aerospace Exploration Agency & Space-Ground VLBI \\
WSRT & Westerbork Synthesis Radio Telescope & Netherlands Institute for & Westerbork, NL \\
 & & Radio Astronomy - ASTRON & \\ \hline
\end{tabular}
\end{center}

\end{table}
\end{landscape}

\begin{landscape}
\begin{table}
\footnotesize
\caption {Subset of those interferometer arrays from Table \ref{table:all_interferometers} which are presently open (or will soon be open) for use by qualified researchers from the general astronomy community. ``$\ast$'' indicates capabilities under development.
\label{table:capabilities}}
\begin{center}
\begin{tabular}{|l|l|l|l|l|l|} \hline
  &  \multicolumn{2}{|c|}{Telescope} & Maximum & Wavelength & Observer Resources \\ 
Acronym &  Number & Size (m) & Baseline  & Coverage & web page link \\  \hline
SUSI &  2 & 0.14 & 64 (640$\ast$) m & 0.5--1.0$\mu$m & http://www.physics.usyd.edu.au/sifa/Main/SUSI \\
NPOI &  6 & 0.12 & 64 ($>250\ast$) m & 0.57--0.85$\mu$m & http://www.lowell.edu/npoi/ \\
CHARA &  6 & 1.0 & 330 m& 0.6--2.4$\mu$m & http://www.chara.gsu.edu/CHARA/ \\
MRO$\ast$ &  $\sim$10 & $\sim$1.5 & $\sim$350 m & 0.6--2.5$\mu$m  & http://www.mro.nmt.edu/Home/index.htm \\
VLTI &  4 & 1.8 -- 8.0 & 130 ($202\ast$) m& 1--2.5$\mu$m, 8--13$\mu$m & http://www.eso.org/sci/facilities/paranal/telescopes/vlti/ \\
Keck-I &  2 & 10.0 & 85 m& 1.5--4$\mu$m, 8--13$\mu$m & Http://planetquest.jpl.nasa.gov/Keck/ \\
LBTI$\ast$ &  2 & 8.4  & 23$\ast$ m& 1--20$\mu$m & http://lbti.as.arizona.edu/ \\
ISI &  3 & 1.65 & 85 ($>100 \ast$) m & 8--12$\mu$m & http://isi.ssl.berkeley.edu/\\
\hline
ALMA$\ast$ & 66 & 7 -- 12  & 15 km  & 0.3 mm -- 3.6 mm & http://science.nrao.edu/alma/index.shtml \\ 
SMA & 8 & 6 & 500 m & 0.4 mm -- 1.7 mm & http://www.cfa.harvard.edu/sma/ \\
CARMA & 23 & 3.5  -- 10.4  & 2000 m &  1.3, 3, 7 mm & http://www.mmarray.org/ \\
PdB & 6 & 15 & 760 m & 1.3, 2, 3 mm & http://www.iram-institute.org/EN/ \\
VLBA & 10 & 25 &  8000~km  &   3 mm -- 28 cm & http://science.nrao.edu/vlba/index.shtml \\
eMERLIN & 7 & 25 -- 76 & 217 km & 13 mm -- 2 m & http://www.e-merlin.ac.uk/ \\
EVN & 27 & 14 -- 305  & $> 10000$ km & 7 mm - 90 cm & http://www.evlbi.org/ \\
VERA & 4 & 20 & 2270 km & 7 mm -- 1.4 cm & http://veraserver.mtk.nao.ac.jp/index.html \\
ATCA & 6 & 22 & 6 km & 3 mm -- 16 cm & http://www.narrabri.atnf.csiro.au/ \\
EVLA  & 27 &  25 & 27 km & 6 mm -- 30 cm & http://science.nrao.edu/evla/index.shtml \\
WSRT  & 14 &  25 &  2700 m &   3.5 cm -- 2.6 m  & http://www.astron.nl/radio-observatory \\
 & & & & & /astronomers/wsrt-astronomers \\
DRAO  &  7  &  9 &  600 m &  21, 74~cm  & e-mail: Tom.Landecker@nrc-cnrc.gc.ca \\
GMRT & 30 & 45 & 25~km & 21 cm -- 7.9 m & http://www.gmrt.ncra.tifr.res.in \\
MOST & 64 & $\sim 12$ & 1600 m & 36 cm & http://www.physics.usyd.edu.au/sifa/Main/MOST \\ 
LOFAR & many & simple & 1500 km & 1.2 m -- 30 m & http://www.astron.nl/radio-observatory \\
 & & & & & /astronomers/lofar-astronomers \\
\hline
\end{tabular}
\end{center}
\end{table}
\end{landscape}

\acknowledgments 

We are grateful to the following colleagues for their comments and
contributions: W.M. Goss, R.D. Ekers, T.L. Wilson, S. Kraus, M. Zhao,
T. ten Brummelaar, A. King, and P. Teuben.

\bibliographystyle{apj}  
\bibliography{apj-jour,pss_monnier_allen_preprint}  

\clearpage


\end{document}